\documentclass[journal=jacsat,manuscript=article]{achemso}

\usepackage{chemformula} % Formula subscripts using \ch{}
\usepackage[T1]{fontenc} % Use modern font encodings
\usepackage{upgreek}
\usepackage{multirow}
\usepackage{booktabs}
\usepackage{subcaption}
\author{Ke Li}
\affiliation{Institute of Theoretical Chemistry, College of Chemistry, Jilin University, 2519 Jiefang Road,
	Changchun 130023, China}
\author{Jitai Yang}
\affiliation{Institute of Theoretical Chemistry, College of Chemistry, Jilin University, 2519 Jiefang Road,
	Changchun 130023, China}
\author{Yu Zhai}
\affiliation{Beijing National Laboratory for Molecular Sciences, Institute of Theoretical and Computational Chemistry, College of Chemistry and Molecular Engineering, Peking University, Beijing 100871, China}
\author{Hui Li}
\affiliation{Institute of Theoretical Chemistry, College of Chemistry, Jilin University, 2519 Jiefang Road,
	Changchun 130023, China}
\email{ prof_huili@jlu.edu.cn}
\date{\today}
\title[An \textsf{achemso} demo]
  {Disentangling Cation-Polyanion Coupling in Solid Electrolytes: Which Anion Motion Dominates Cation Transport?}
\begin{document}

%\begin{tocentry}

%Some journals require a graphical entry for the Table of Contents.
%This should be laid out ``print ready'' so that the sizing of the

%Inside the \texttt{tocentry} environment, the font used is Helvetica
%%8\,pt, as required by \emph{Journal of the American Chemical

%The surrounding frame is 9\,cm by 3.5\,cm, which is the maximum
%permitted for  \emph{Journal of the American Chemical Society}
%graphical table of content entries. The box will not resize if the
%content is too big: instead it will overflow the edge of the box.

%This box and the associated title will always be printed on a
%separate page at the end of the document.

%\end{tocentry}

\begin{abstract}
Lithium and sodium solid electrolytes feature polyanion frameworks and highly mobile cations. Understanding and quantifying the impact of polyanion dynamics on cations will help us to unravel the complex role that anion play in superionic conductors. %unravel the complex role that anion dynamics play in superionic conductors. %Constituting highly conductive inorganic solid electrolytes, has attracted significant research interest, yet its complex dynamics that govern fast ion conduction remain largely elusive. 
However, no experimental or computational method can directly extract this information, as polyanion dynamics are always coupled with other factors that affect ion mobility.
Here, we present the pioneering study that combines constraint algorithm and machine-learning molecular dynamics to quantitatively reveal the effects of polyanion translation, rotation, and vibration on cation mobility across a diverse material class. %A diverse selection of inorganic superionic conductors from multiple distinct classes was chosen to ensure the universality of the results. 
Ultralong-time, large-scale machine-learning molecular dynamics simulations with selective constraints on each anion motion mode unequivocally yield results at near room and elevated temperatures. 
In sharp contrast to the previous understanding that facile anion rotation primarily facilitates cation transport, the strong coupling between anion translation and vibration with cation diffusion has been unraveled for the first time; we find that translation, rotation, and vibration can each directly drive superionicity, with one typically dominant in each class of materials.
Anion rotation dominates cation transport when its frequency matches the cation hopping frequency, whereas anion translation prevails at higher and vibration at lower rotation frequencies.
The impact of anion dynamics on cation diffusion becomes more prominent at lower temperatures.
Further in-depth analyses uncover that, fundamentally, these anion motion modes drive cation migration by inducing large disorder within the cation sublattice. In addition, anion dynamics achieve enhanced ion mobility both by introducing fluctuations in the underlying energy landscape, thereby effectively lowering the energy barrier, and by altering local cation dynamics, which directly affects the prefactor. These findings demonstrate that polyanion dynamics are at the origin of superionicity and provide principles for the future discovery and development of advanced superionic conductors.
  %This is an example document for the \textsf{achemso} document
  %class, intended for submissions to the American Chemical Society
  %for publication. The class is based on the standard \LaTeXe\
  %\textsf{report} file, and does not seek to reproduce the appearance
  %of a published paper.

  %This is an abstract for the \textsf{achemso} document class
  %demonstration document.  An abstract is only allowed for certain
  %manuscript types.  The selection of \texttt{journal} and
  %\texttt{manuscript} will determine if an abstract is valid.  If
  %not, the class will issue an appropriate error.

%soh, nobody ever pays much attention to the dynimical factor before.
\end{abstract}


\section{Introduction} 
All-solid-state battery, which offers high energy and improved safety, has received considerable attention and extensive research in recent years\cite{li2022long,ma2023high}. One of its key components, the solid electrolyte (SE), is crucial to the device's overall performance and much effort has been contributed to searching for SE with high ionic conductivity. The rational design of advanced conductive materials largely hinges on a fundamental understanding of its nature at the atomic level\cite{zou2021identifying}. 

From a static perspective, thermally activated cation hopping occurs within rigid potential wells defined by the host lattice, allowing cations to move from site to site by overcoming migration energy barriers. Specific anion packing\cite{wang_design_2015} and anion site disorder\cite{rayavarapu2012variation,kraft_inducing_2018} can facilitate cation diffusion. In addition to these factors, dynamic mechanisms have been proposed, including lone pair--ion dynamics\cite{mercadier_dynamic_2023,dhattarwal_electronic_2024} and the cooperative motion of the host lattice framework and charge carriers\cite{paulus_lattice_2008,lei_dynamic_2024}. Among these mechanisms, dynamic disorder arising from polyanion motion has been extensively observed, with its rotational disorder receiving significant attention\cite{zhang_coupled_2019,sun_rotational_2019,zhang_targeting_2020, maughan_lowering_2021,scholz_superionic_2022,tsai_double_2023}. 
Rotational disorder is commonly observed in various inorganic superionic conductors, including sulfates\cite{lunden_enhancement_1988}, phosphates\cite{witschas_anion_2000,saha_structural_2020}, hydrides\cite{matsuo_lithium_2007}, closo-borates\cite{udovic_sodium_2014,tang_unparalleled_2015,udovic_exceptional_2014,tang_liquidlike_2016,sadikin_modified_2017}, antiperovskites\cite{sun_rotational_2019,tsai_double_2023}, and thiophosphates\cite{zhang_coupled_2019,scholz_superionic_2022,famprikis_new_2019,zhang_targeting_2020}.
A closely related and highly debated concept is the paddle-wheel effect, first proposed over thirty years ago\cite{jansen_volume_1991,zhang_exploiting_2022}. This effect suggests that the rotational motion of polyanions enhances cation migration. While the superionic behavior of cations is believed to be coupled with anion rotation, and recent studies have leveraged this principle in the search for superionic conductors\cite{yang_harvest_2024}, the existence of the paddle-wheel effect remains controversial and unclear. Recently, this effect has been subjected to renewed scrutiny\cite{kweon_structural_2017,zhang_coupled_2019,smith_low-temperature_2020,sun2022enhanced,forrester_disentangling_2022,xu_machine_2023,jun_nonexistence_2024,gigli_mechanism_2024}.

Several representative materials have laid the foundation and provided templates for designing novel materials that exploit the paddle-wheel effect. In 2007, Matsuo et al. \cite{matsuo_lithium_2007} discovered that the ionic conductivity of lithium borohydride at $\sim$390\:K surged dramatically by three orders of magnitude to 
\(10^{-3}\:\text{S\:cm}^{-1}\). This increase was accompanied by a structural transition from the orthorhombic phase to the hexagonal phase and the onset of quasi-free rotation of \ch{BH4^{-}} anion., 
Due to its relatively low transition temperature, stabilizing the metastable polymorph---and thus achieving superionic conductivity---at room temperature is very promising. Indeed, incorporating lithium halide into the system or nanoconfining it can stabilize the high-temperature phase and maintain the anion rotation near/at room temperature, resulting in fast \ch{Li^{+}} conduction.\cite{maekawa_halide-stabilized_2009,blanchard_nanoconfined_2015}. However, although much work has focused on utilizing the superionic hexagonal phase, the mechanism behind the sudden increase in conductivity in the host material, \ch{LiBH4}, and the relationship of this increase to anion rotations (and possibly other motions) are still not fully understood.
In the dehydrogenation process of \ch{LiBH4}, an intermediate product, \ch{Li2B12H12}, is formed and exhibits high conductivity in its superionic phase. The related electrolyte family spans an extremely wide compositional and configurational space, including closo-type compounds\cite{udovic_sodium_2014,tang_unparalleled_2015,udovic_exceptional_2014,tang_liquidlike_2016,sadikin_modified_2017} (e.g. \ch{M2B}$_n$\ch{H}$_n$ ($n$=10,12), \ch{MCB}$_n$\ch{H}$_n$ ($n$=9,11), M=Li, Na; \ch{Na2B12H}$_{12-x}$\ch{I}$_x$), nido-type compounds \cite{tang_orderdisorder_2017} (e.g. \ch{NaB11H14}, \mbox{\ch{Na}-7-\ch{CB10H13}}, \mbox{\ch{Li}-7-\ch{CB10H13}}, \mbox{\ch{Na}-7,8-\ch{C2B9H12}}, and \mbox{\ch{Na}-7,9-\ch{C2B9H12}}). These materials exhibit remarkable electrical conductivity, reaching up to 70\:mS\:cm$^{-1}$ through anion mixing, even at ambient temperature\cite{tang_stabilizing_2016}. Studies have shown that, in this class of materials, the orientational disorder of anions is linked to fast cation transport\cite{kweon_structural_2017,lu_structural_2016,varley_understanding_2017,sau_comparative_2021}.
Another newly developed sodium-based SE, \ch{Na3OBH4}, exhibits conductivity comparable to that of liquid counterpart. %owing in part to the paddle-wheel effect\cite{sun_rotational_2019}. 
This antiperovskite-type ionic conductor features a dual-anion structure and a highly rotationally mobile \ch{BH4^{-}}. The rotation of \ch{BH4^{-}} anions is proposed to facilitate \ch{Na^{+}}-ion transport, enhancing ionic mobility within the lattice\cite{sun_rotational_2019,zhao_design_2022,tsai_double_2023}.

Lithium thiophosphate represents another promising family of electrolytes and is derived from the binary \ch{Li2S}--\ch{P2S5} system, with distinct structural units such as \ch{PS4^{3-}} and \ch{P2S6^{4-}}, depending on composition\cite{kudu_review_2018}. One archetypal system, \ch{Li3PS4}, constituted of ortho-thiophosphate units (\ch{PS4^{3-}}), transitions from the $\gamma$ to $\beta$ phase at intermediate temperature, along with an anomalous increase in conductivity and the onset of anion rotation\cite{homma_crystal_2011,zhang_targeting_2020}. The $\beta$ polymorph continues to undergo phase transition to the $\alpha$ polymorph at higher temperature, which was proposed to show exceptional conductivity\cite{kim_thermally_2018}. Efforts to stabilize these two highly conductive phases at room temperature have included partial substitution of Si at the P site in the $\beta$ rotor phase\cite{zhang_targeting_2020} and rapid heating of glass to obtain the stabilized $\alpha$ phase\cite{kimura_stabilizing_2023}. However, great controversy has arisen regarding the existence of the paddle-wheel effect in this type of compound.\cite{xu_machine_2023,smith_low-temperature_2020,jun_nonexistence_2024,gigli_mechanism_2024,forrester_disentangling_2022,zhang_targeting_2020}
Indeed, the \ch{PS4^{3-}} anion appears in many systems known for record-high conductivity,
\cite{seino_sulphide_2014,kamaya_lithium_2011,zhang_na11_2018,deiseroth_li6_2008,famprikis_new_2019} but the coupling between its dynamics and cation motion remains poorly understood. A novel sodium-based superionic conductor, \ch{Na11Sn2PS12}, is comprised of a characteristic three-dimensional chessboard framework of \ch{PS4} and \ch{SnS4} tetrahedra and achieves very high room-temperature conductivity\cite{zhang_na11_2018}. In contrast to other systems, which transition to the plastic phase at elevated temperatures, this material maintains its phase integrity and anion rotation, which was proposed to aid cation diffusion, down to very low temperatures.\cite{zhang_coupled_2019}. 
Another recently discovered sodium thiophosphate electrolyte, \ch{Na4P2S6}, exhibits high phase-space flexibility and excellent conductivity in its high-temperature 
$\gamma$ phase\cite{scholz_superionic_2022}. In the $\gamma$ phase, orientational disorder of the prolate anion \ch{P2S6^{4-}}, which has relatively low point-group symmetry compared to other common rotors, and flexibility of the anion dihedral angle are suggested by both experimental and computational studies. The dynamic coupling between liquidlike sodium conduction and different modes of motion of this ethan-like anion remains elusive too. %, is far from being fully understood.

Molecular dynamics (MD) can detect the paddle-wheel effect by probing local dynamics, identifying cooperative anion rotation and cation migration events that occur in close temporal and spatial proximity \cite{zhang_coupled_2019,jun_nonexistence_2024,xu_machine_2023,smith_low-temperature_2020}. However, these observations cannot fully quantify the overall changes in cation diffusion properties resulting from anion rotation. Additionally, rotational motions typically manifest following phase transitions, making it experimentally challenging to isolate their effects from other variables. Furthermore, while anion rotation receives considerable attention, other degrees of freedom—such as the translation and vibration of polyanions—are often overlooked, and there are few studies examining their effects on cation migration. Machine learning is a powerful technique for extending the spatiotemporal scales of simulations, effectively balancing the trade-off between simulation speed and accuracy\cite{sahrmann2023utilizing,huang_deep_2021,lin2021unravelling}.

Lattice dynamics, proposed as a descriptor long ago \cite{mahan_lattice_1976}, is now emerging as a key strategy for tuning ionic mobility \cite{muy_phononion_2021}. Ion migration within crystals is closely related to phonon; therefore, understanding lattice vibrations is of paramount importance. Several studies have sought to relate diffusive properties to acoustic phonons. K\"{o}hler et al. correlated atomic mobility with longitudinal acoustic phonons at 2/3<111>, finding that lower phonon frequencies correspond to lower activation energies\cite{kohler_correlation_1988}. Kraft et al. linked the Debye frequency—estimated from measurements of the speed of sound associated with the acoustic branch—to migration enthalpy and the prefactor \cite{kraft_influence_2017}. In terms of optical phonon, Wakamura proposed a simple model demonstrating that activation energy scales quadratically with the frequency of low-energy optical modes \cite{wakamura_roles_1997}. Krauskopf et al. employed Raman spectroscopy to investigate the correlation between optical modes and ion migration processes, discovering that a red shift in optical modes leads to lower migration enthalpy \cite{krauskopf_comparing_2018}. Typically, the translation and rotation of polyanions occur within the low-frequency range, while higher-frequency optical modes are dominated by intramolecular vibrations of the polyanions. However, the mechanisms by which these phonon modes contribute to fast ion conduction, as well as the extent of their influence, remain poorly understood.

In this study, using our newly developed constraint algorithm\cite{yang_rotationalroto-translational_2024}, we systematically estimate the quantitative contributions of anion translation, rotation, and vibration to conductivity and, more importantly, unveil the mechanisms by which anion dynamics enable superionicity. We performed extensive, ultralong machine learning molecular dynamics (MLMD) simulations to ensure access to reliable results at near room temperature and at elevated temperatures. Utilizing six superionic conductors from three distinct material classes, we have identified distinct anion motion modes that play a critical role in influencing conductivity. Furthermore, we demonstrate that anion dynamics can induce cation disorder and lower activation energy. We hope this study can resolve previous controversies and provide robust support for the growing body of research that leverages dynamic coupling as a design guideline \cite{zhang_exploiting_2022,yang_harvest_2024,tsai_double_2023}, thereby advancing battery design development.
\section{Results and discussion}
\subsection{Anion Rotation and Constraint Algorithm}
The crystal structures of the six studied materials, obtained from previous studies\cite{udovic_exceptional_2014,soulie_lithium_2002,sun_rotational_2019,kaup_impact_2020,zhang_na11_2018,scholz_superionic_2022}, are shown in Figure~1. In these unique plastic crystal structures, polyanion reorientations and partial occupancy of mobile cation are observed. The machine learning interatomic potentials (MLIPs) were trained on these plastic phases. 
The root mean square errors (RMSE) for energies and forces of the models for each system are listed in Table~1. The RMSE for energies is around 1\:meV per atom, and for forces, it is several tens of meV/\AA, demonstrating a high level of fitting accuracy. The low standard deviations in both energies and forces indicate uniform accuracy of the each model ensemble. Meanwhile, a small disparity between test and training data suggests no explicit overfitting. The facile reorientation of anions was well reproduced (see below). Diagnostic plots and training details are provided in Figure~S1--S2 and Section~1 of the Supporting Informations.
\begin{figure}[h]
\centering
\includegraphics[width=\textwidth]{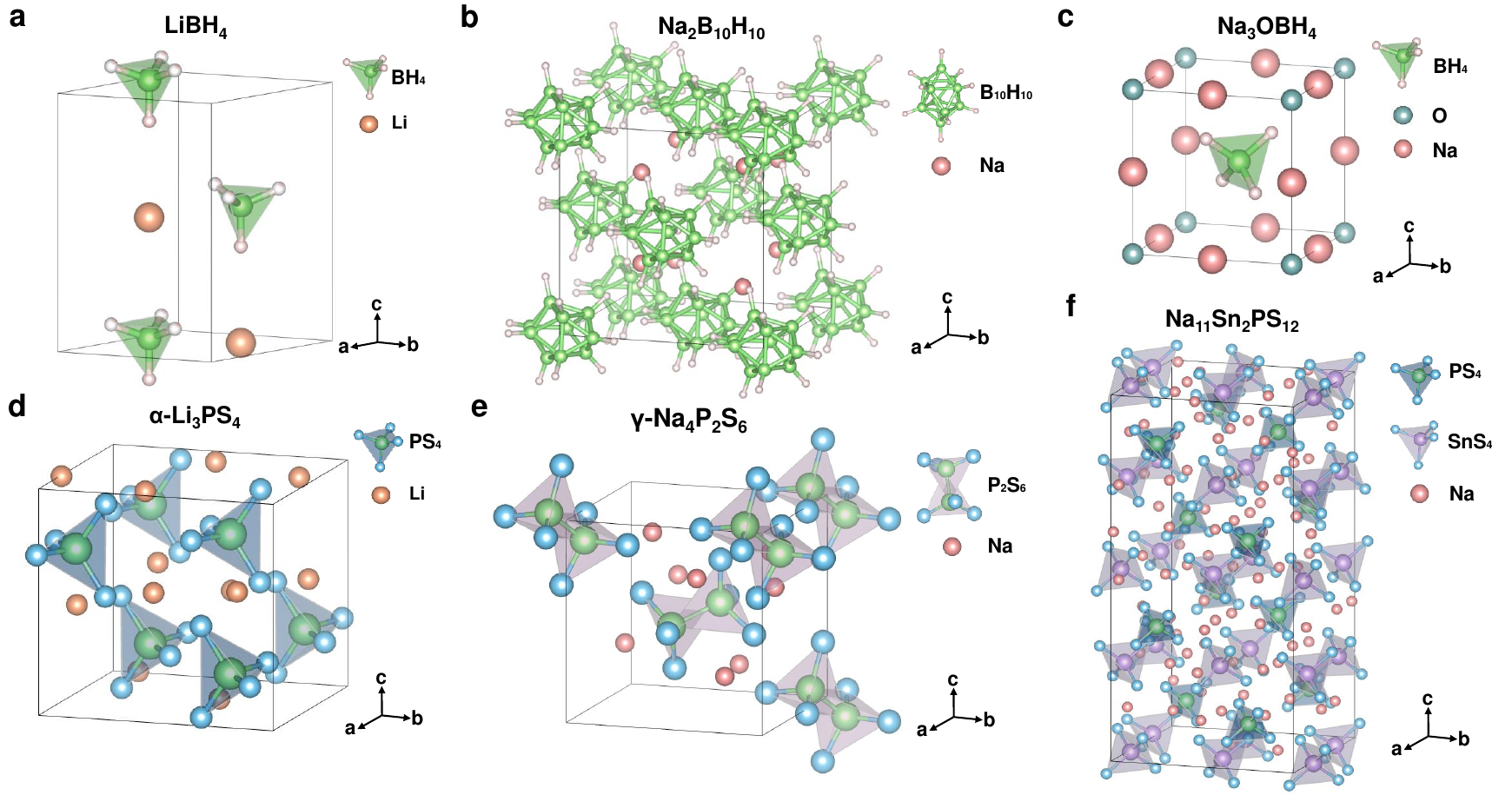}
\caption*{\textbf{Figure\:1.}~(a) High-temperature hexagonal polymorph of \ch{LiBH4} in \textit{P}6\textsubscript{3}\textit{mc} space group. (b) Unit cell of liquid-like phase of \ch{Na2B10H10} with \textit{Fm\(\bar{3}\)m} symmetry. (c) Antiperovskite \ch{Na3OBH4} with cubic structure with \textit{Pm\(\bar{3}\)m} symmetry. (d) The $\alpha$-polymorph of \ch{Li3PS4} crystalized in \textit{Cmcm} space group. (e) $\gamma$-modification of \ch{Na4P2S6} with \textit{Im\(\bar{3}\)m} symmetry. (f) Crystal structure of \ch{Na11Sn2PS12} with tetragonal \textit{I}4\textsubscript{1}/\text{acd} symmetry.}
\label{fig1}
\end{figure}

\setlength{\tabcolsep}{10pt}
\def\arraystretch{1.1}%
\begin{table}
   \caption*{Table 1. RMSE of energies per atom (meV/atom) and force (meV/\AA) for DP models of six superionic conductors. The standard deviations are evaluated from the ensembles of four models.}
\begin{tabular}{@{}lllll@{}}
\toprule
Materials & \multicolumn{2}{l}{RMSE in energy}  & \multicolumn{2}{l}{RMSE in force}\\ \cmidrule{2-5}
& Training & Test & Training & Test\\ 
\midrule
\ch{Na11Sn2PS12} &   {0.60 \(\pm\) 0.09}        & 0.71 \(\pm\) 0.01   & 43.1 \(\pm\) 5.70   & 41.8 \(\pm\) 0.26    \\ 
\ch{Na4P2S6} & {0.37 \(\pm\) 0.03}        & 1.1 \(\pm\) 0.03   & 49.8 \(\pm\) 0.95   & 52.4 \(\pm\) 0.54    \\ 
\ch{Na2B10H10} & {1.03 \(\pm\) 0.03}        & 1.31 \(\pm\) 0.18   & 79.3 \(\pm\) 1.51   & 56.4 \(\pm\) 0.92    \\ 
\ch{Na3OBH4} & {0.42 \(\pm\) 0.06}        & 0.50 \(\pm\) 0.01   & 28.4 \(\pm\) 0.72   & 12.0 \(\pm\) 0.30    \\ 
\ch{LiBH4} & {1.22 \(\pm\) 0.03}        & 1.74 \(\pm\) 0.13   & 57.1 \(\pm\) 0.35   & 35.6 \(\pm\) 0.27    \\ 
\ch{Li3PS4} & {1.44 \(\pm\) 0.01}        & 1.10 \(\pm\) 0.04   & 89.3 \(\pm\) 0.61   & 71.1 \(\pm\) 0.60    \\ 
\bottomrule
\end{tabular}
\label{TableDP}
\end{table}
We performed MLMD simulations with no constraint (UC), rotational constraint (RC), roto-translational constraint (RTC), and roto-translational-vibrational constraint (RTVC) on orientationally disordered anions at various temperatures. For RC and RTC simulations, we explicitly constrain each rotor anion's 3 rotational degrees of freedom (DOF) and 6 translational and rotational DOF, respectively. In RTVC simulations, we constrain all DOF of the each rotor anion. The angles (\(\theta_{x}\), \(\theta_{y}\) and \(\theta_{z}\)) between the orientation vector (the red dotted line) and three axes (parallel to the crystal lattice reference frame) are defined for each type of rotor in the insets of Figure~2a--c, respectively, to describe the evolution of anion reorientations over time.
The results are shown in Figure~S3. Note that in \ch{Na11Sn2PS12}, the \ch{SnS4^{4-}} typically does not exhibit any rotational mobility\cite{zhang_coupled_2019}; therefore, only \ch{PS4^{3-}}, which displays orientational disorder, is considered. In addition, transitions between different staggered conformations were observed for \ch{P2S6^{4-}} in \ch{Na4P2S6} (Figure~S4), but typically at frequencies below 0.05\:ns$^{-1}$ per rotor. We observe that anions composed of the same elements with similar point-group symmetry exhibit comparable rotation frequencies, and their rotation frequencies show similar temperature-dependent susceptibility, regardless of variations in lattice symmetries and chemical environments, as illustrated in Figure~2e. Based on this, we classify the rotors into three types, and the representative angle projections are shown in Figure~2a--c. The slow rotors, \ch{PS4^{3-}} and \ch{P2S6^{4-}} anion groups, become orientationally disordered only at elevated temperatures (approximately 900\:K) and exhibit slow transitions among preferred orientations. For the \ch{BH4^{-}} rotor in lithium hydride and antiperovskite, the rotations are relatively free, and its rotation frequency is highly susceptible to temperature changes. This difference likely arises from the larger moments of inertia of \ch{PS4^{3-}} and \ch{P2S6^{4-}} compared to \ch{BH4^{-}}, as well as differences in interaction strength between these rotor anions and the cation framework. The last type of rotor is \ch{B10H10^{2-}} with high point-group symmetry  \textit{D}\textsubscript{4\textit{d}}. Despite its large moment of inertia, the high intrinsic symmetry lowers the energy barrier for fast reorientational motion, resulting in a moderate rotation frequency. Free energy surfaces for the rotation of three types of rotors of representative materials show a consistent trend (Figure~S5). Figure~2d shows the angle projection of representative \ch{BH4^-} rotor with RC applied. The absence of large angle fluctuation indicates the successful implementation of the rotation constraint algorithm.

We performed a statistical ion hopping analysis on each superionic conductor under UC, and the results, along with the corresponding rotor anion rotation frequencies, are presented in Figure~2f.  All thiophosphates exhibit cation jump frequencies that exceed their anion orientation transition frequencies. This difference is particularly pronounced in \ch{Li3PS4}, where the cation jump frequency is approximately four orders of magnitude higher than the anion orientation transition frequency. In \ch{LiBH4}, despite the presence of fast rotational dynamics of the \ch{BH4^{-}} rotor, the \ch{Li^{+}} jump frequency is significantly lower. Interestingly, one of the simulated electrolytes, \ch{Na3OBH4}, with the same fast \ch{BH4^-} rotors constituting the sublattice, showed no successful \ch{Na^+} hops after simulation at 800\:K for 10\:ns within a 1125-atom box (see the mean square displacment (MSD) result in Figure~S6), contradicting experimental findings \cite{sun_rotational_2019}. Ahiavi et al.\cite{ahiavi_mechanochemical_2020} and Tsai et al. \cite{tsai_double_2023} suggest that the high conductivity cannot be reproduced neither by experiment nor MD simulation. A comparable match between the two frequencies is observed only in \ch{Na2B10H10}, as indicated by a previous study\cite{dimitrievska_carbon_2018}. It has been suggested that for the paddle-wheel mechanism to make a significant contribution to cation hopping, a similar event frequency between rotational motion and translational motion would be required\cite{zhang_exploiting_2022,jun_nonexistence_2024}. The mismatches between anion reorientational frequencies and cation jump frequencies contradict the hypothesis that large-angle anion rotations directly facilitate cation hopping through the docking-undocking process\cite{kweon_structural_2017}. %This raises the question of whether the plastic nature facilitates cation diffusion.
 %It can also be seen that the angle fluctuation amplitude of the constrained case is smaller than the unconstrained case, demonstrating a deprivation of the libration motion. 
\begin{figure}[H]
\centering
\includegraphics[width=\textwidth]{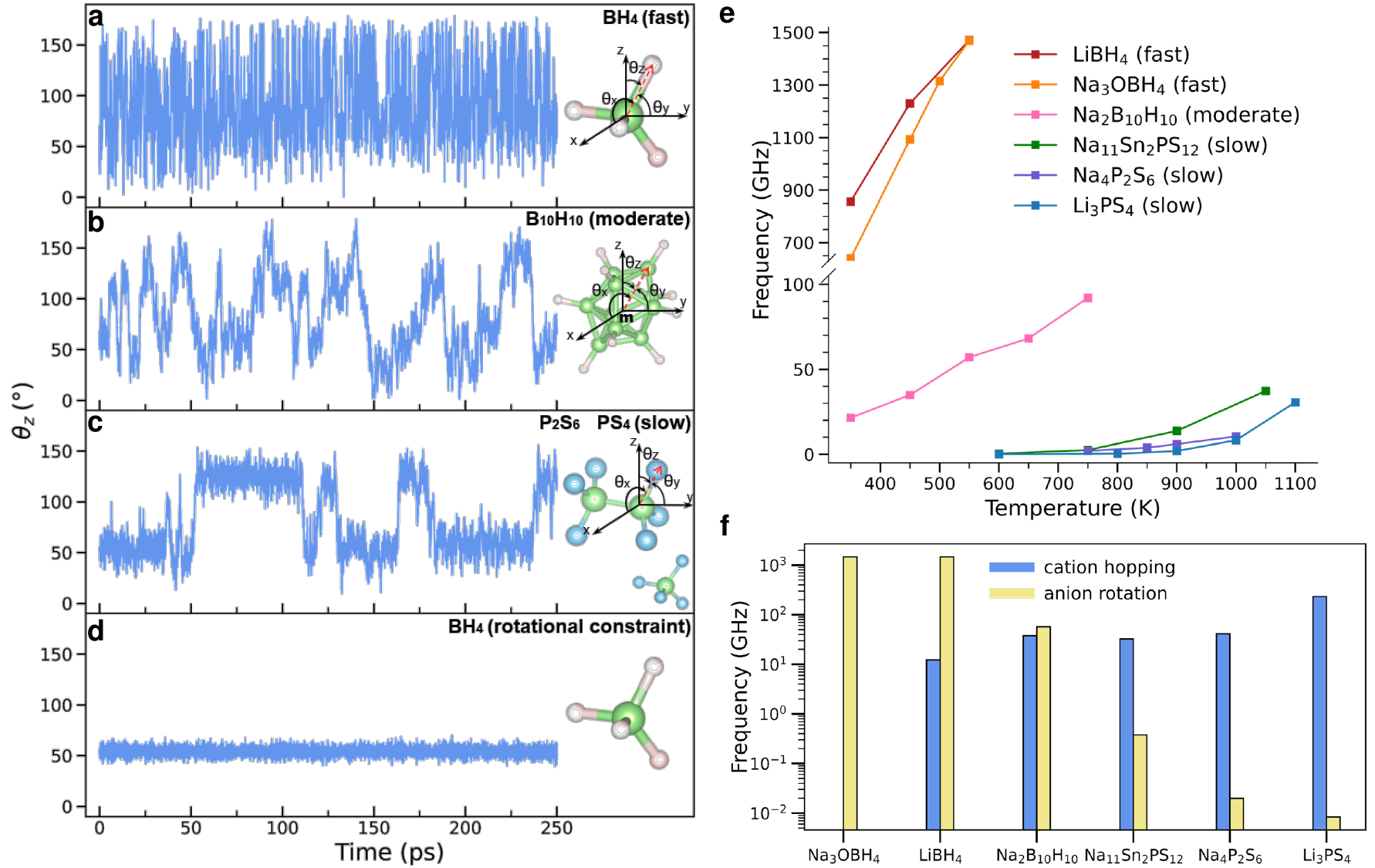}
\caption*{\textbf{Figure\:2.}~Angle projections of (a) \ch{BH4^-} in \ch{LiBH4} at 500\:K, (b) \ch{B10H10^{2-}} in \ch{Na2B10H10} at 650\:K, and (c) \ch{PS4^{3-}} in \ch{Na11Sn2PS12} at 1050\:K, representing fast, moderate, and slow reorientational dynamics, respectively. The internal coordinates for rotors with different symmetries are defined, as shown in the insets. The \textbf{m} denotes the center of mass of anion. (d) Representative angle projection of \ch{BH4^{-}} with the rotational constraint algorithm applied. The small fluctuations arise from vibrational wiggling of the atoms. (e) Extracted reorientational frequencies against simulation temperatures. Angles fluctuations greater than 70$^\circ$ are considered successful rotations. These frequencies align well with the reorientation frequencies measured in previous experiments and those simulated using $ab$ $initio$ molecular dynamics (see  section\:1 in Supporting Information). (f) Cation hopping frequencies and anion reorientation frequencies for six ionic conductors. The simulated temperature is 550\:K for the borane and antiperovskite compounds, and 600\:K for the thiophosphates. The frequencies are normalied by the number of cations and rotors, respectively. For \ch{Na3OBH4}, \ch{Na^+} showed no successful hops in our simulation.}
\label{fig1}
\end{figure}
\subsection{Quantifying Cation-Anion Coupling Dynamics}
To quantify how different types of anion framework with distinct reorientational mobility dynamically affect cation transport, we calculated the conductivity under UC, RC, RTC, and RTVC of each conductor at various temperatures, as comparatively shown in Figure~3a--e.  In some cases, no net jump events occurred, or only limited ion diffusion was captured, with a total mean-squared displacement (MSD) of just a few Angstrom within 10\:ns; calculating reliable values for conductivity was not feasible. Their MSD, shown in Figure~S7--S10, fluctuate around a constant value for a long period. Universally, anion dynamics facilitate the diffusion of cations, leading to higher conductivities. In lithium borohydride and sodium decahydro-closo-decaborate, with the same elements constituting the anion sublattices, cation mobility exhibits stronger responses to anion dynamics the thiophosphate.
More importantly, the contributions of anion translation, rotation, and vibration to conductivity in each superionic conductor varies, with one consistently dominating across all simulated temperatures (Figure~S11--S12). 

For \ch{Na2B10H10}, anion rotation dominantly drives cation diffusion across the entire simulated temperature range (Figure~3a). Applying the rotational constraint reduces conductivity by at least two orders of magnitude. The absence of translation further reduces conductivity by approximately one order of magnitude at elevated temperatures. In the high-temperature polymorph of \ch{LiBH4}, the \ch{BH4^{-}} tetrahedron undergoes rapid quasi-free rotation, which is also crucial for \ch{Li^{+}} ion diffusion across a wide range of SE operating temperatures (Figure~3b). Applying the rotational constraints to the \ch{BH4^{-}} anion matrix results in an order of magnitude decrease in conductivity. Surprisingly, no \ch{Li^{+}} ion hop events were observed when the RTC was applied at all simulated temperatures, each simulating 10\:ns. The absence of \ch{BH4^-} translation leads to a decrease in conductivity of more than three orders of magnitude at elevated temperatures. This indicates that \ch{LiBH4} transitions from superionic conductor to an ionic `insulator' within the simulated time and system scale, highlighting the strong dynamic coupling between the translational motion of the \ch{BH4^{-}} and \ch{Li^{+}} diffusion. MD simulations of \ch{LiBH4} with \ch{BH4^-} under RTVC were also conducted. The conductivity cannot be calculated in this case due to drastic decrease in conductivity resulting from the absence of anion translation; therefore, the contribution of \ch{BH4^-} vibration cannot be estimated. 
The rotation of \ch{B10H10^{2-}} in \ch{Na2B10H10} and the translation of \ch{BH4^-} in \ch{LiBH4} are the key modes enabling superionicity, as they are the primary causes of fast diffusion.
\begin{figure}[H]
\centering
\includegraphics[width=\textwidth]{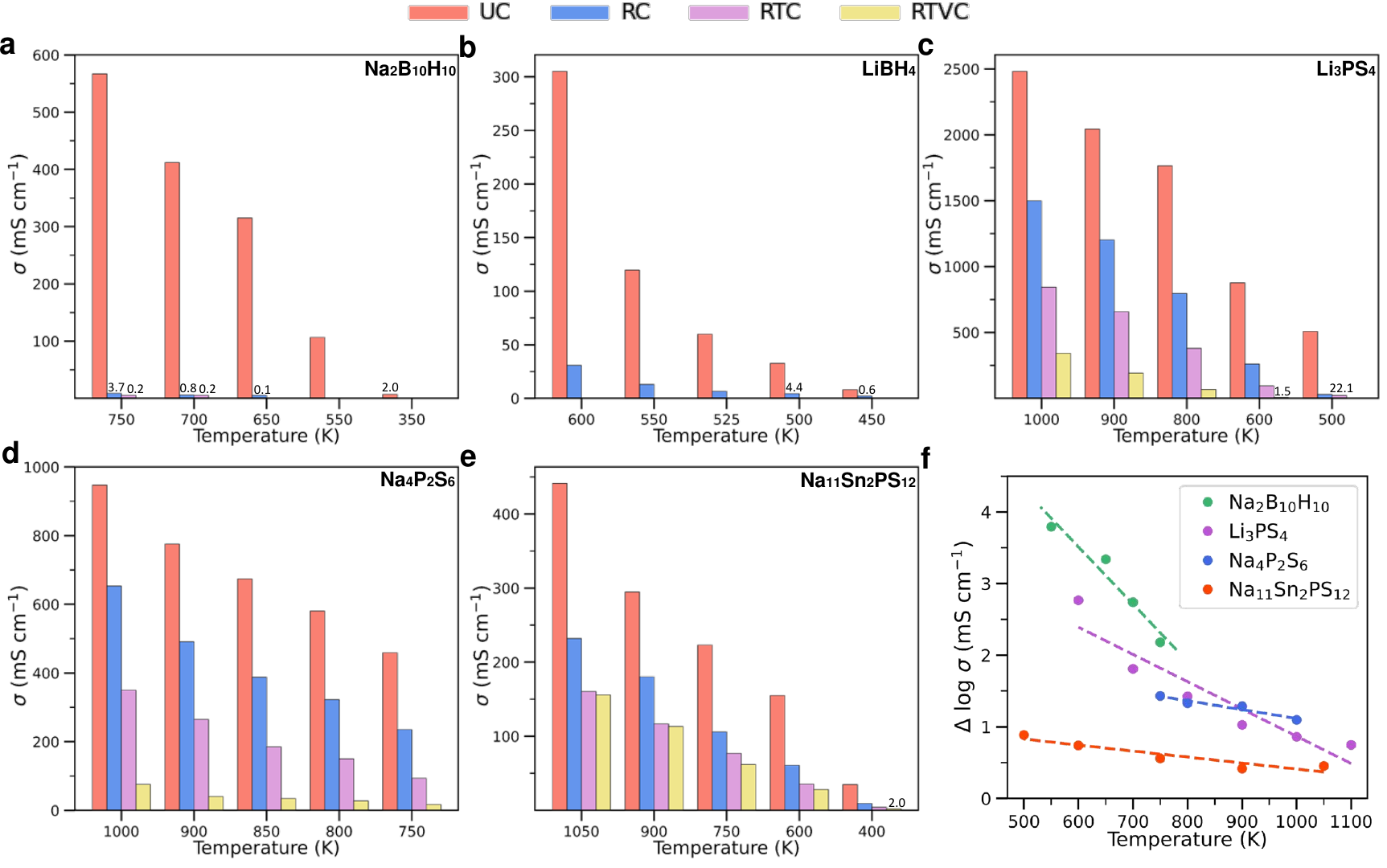}
\caption*{\textbf{Figure\:3.}~Conductivity of five materials: (a) \ch{Na2B10H10}, (b) \ch{LiBH4}, (c) $\alpha$-\ch{Li3PS4}, (d) $\gamma$-\ch{Na4P2S6}, and (e) \ch{Na11Sn2PS12} under four constrained conditions at various temperatures. Some conductivities are absent due to the failure to obtain results within the 10\:ns simulations. (f) Temperature versus order-of-magnitude decreases in conductivity due to the application of constraints on rotor anions. For thiophosphates, the decrease in conductivity was calculated under RTVC. For \ch{Na2B10H10}, only the contribution of rotation was considered, as it accounts for the primary decrease in conductivity. For \ch{LiBH4}, the absence of \ch{BH4^-} translation reduces conductivity below the measurable range within the simulated timescale, leaving its temperature dependence undefined.}
\label{fig3}
\end{figure}
For thiophosphate-type SEs, the dynamic coupling between cations and anions remains a key factor in enhancing ionic conductivity, with all motion modes facilitating cation diffusion (Figure~3c--e). However, the conductivity generally stays above $1$\:mS\:cm$^{-1}$, even after applying RTVC. This suggests that the highly polarizable anion sublattice and the relatively weak bonding interactions between cations and anions are also important factors contributing to superionicity in this class of materials\cite{kraft_influence_2017,wang_design_2015,muy_phononion_2021}. In thiophosphates, cation migration is sensitive to the translation and rotation of anions, but exhibits varying sensitivity to their vibrational degrees of freedom. The vibrations of \ch{PS4^{3-}} polyanion have a negligible effect on \ch{Na^+} migration in \ch{Na11Sn2PS12}, whereas in another sodium conductor, \ch{Na4P2S6}, the vibrations of the \ch{P2S6^{4-}} polyanion strongly influence cation migration. Anion vibration is also crucial in \ch{Li3PS4}. At 500\:K under RTVC, the conductivity decreases from 500\:mS\:cm$^{-1}$ to below the measurable range (0.1\:mS\:cm$^{-1}$), representing a drop of more than three orders of magnitude caused by the vibrational constraint. This indicates the strong dynamic coupling between \ch{Li^+} motion and \ch{PS4^{3-}} vibration at lower temperatures. 
Several studies have performed MD simulations with the anion sublattice fully fixed to evaluate the effect of anion rotation on cation migration\cite{ kweon_structural_2017,tsai_double_2023}. However, this approach is ambiguous, as it cannot attribute the effect to a specific motion mode and may even lead to an incorrect physical interpretation. For instance, \ch{BH4^-} translation dominates Li diffusion in \ch{LiBH4}, despite its fast rotation. Here, we identify the dominant anion motion mode that most significantly affects conductivity in each superionic conductor.
The temperature dependence of the order-of-magnitude decrease in conductivity caused by dynamic couplings is shown in Figure~3f. These anion dynamics are crucial for enabling superionicity, and become even more critical as the temperature decreases.
\subsection{Correlating Anion Dynamics with Migration Enthalpy and Prefactor}
To gain a deeper insight into the mechanism underlying dynamic cation-anion coupling, we calculated the activation energy (Figure~4a) and free energy (Figure~4b and Figure~S13) of cation migration under various constraints, and the Arrhenius plots can be found in Figure~S14. The conductivity (or diffusion coefficient) reflects the collective ion diffusion characterized by Brownian motion, whereas ion migration in solids can be described by absolute rate theory\cite{vineyard_frequency_1957}, where activation energy governs both the formation and migration of mobile ions. The results indicate that anion rotation (or libration) lowers the energy barrier, in line with previous proposals that anion rotation introduces fluctuations in the local potential energy landscape and reduces the effective energy barrier\cite{zhang_coupled_2019,kweon_structural_2017,varley_understanding_2017}. In addition, we find that other degrees of freedom can also lower the energy barrier and even emerge as the dominant factor (e.g. the \ch{PS4^{3-}} vibration in \ch{Li3PS4}). Note that the quantitative accuracy of the barriers should not be overemphasized; but the trends should be robustly predicted, with an increase in barriers corresponding to a decrease in conductivity. 
\begin{figure}[H]
\centering
\includegraphics[width=\textwidth]{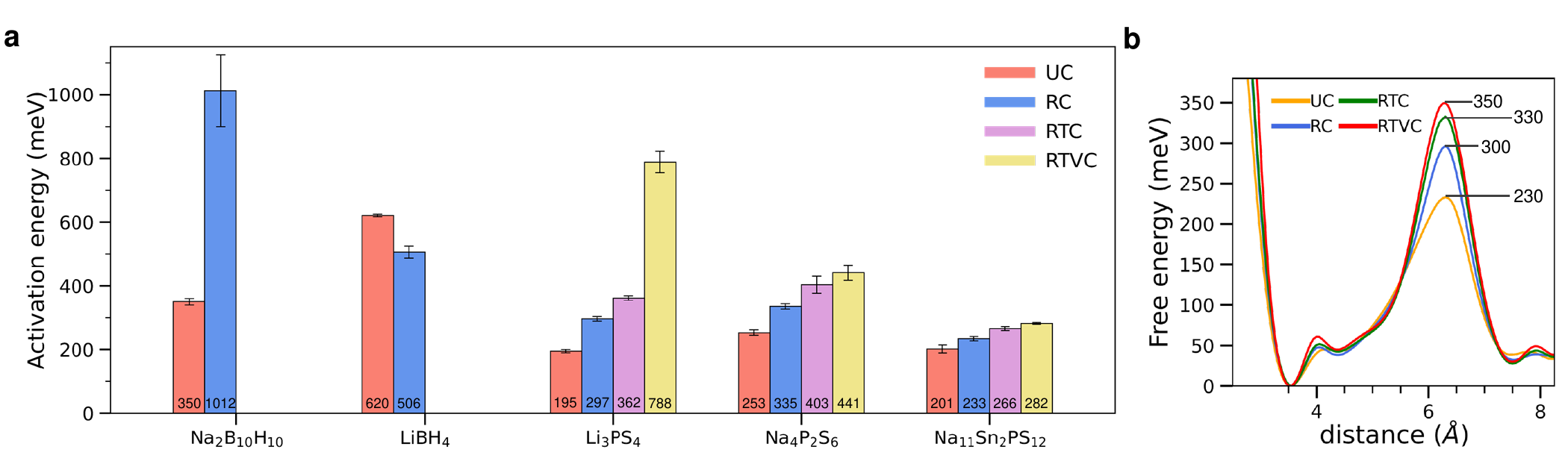}
\caption*{\textbf{Figure\:4.}~(a) Activation energies derived from linear Arrhenius fits of five superionic conductors under various constrained cases. For the two boranes under RTC and RTVC, the activation energies are too high, and the diffusion coefficients over a wide temperature range cannot be obtained; therefore, the activation energies are absent. (b) Free energy curves extracted from \ch{Na11Sn2PS12} trajectory at 900\:K; the distance is between cation and center of mass of orientationally disordered anion.}
\label{fig4}
\end{figure}
However, an exception is observed when the fast rotation of \ch{BH4^-} in \ch{LiBH4} is constrained, reducing the conductivity by about one order of magnitude (Figure~3b), with a simultaneous lowering of activation energy. This suggests a decrease in the prefactor due to the rotational constraint. Fitting the prefactor from MD simulations is not feasible due to significant errors, but a closer examination of the parameters influencing the prefactor still yields qualitative insights. The prefactor is expressed as\cite{tilley_defects_2008} 
\begin{equation}
\sigma_{0}=\frac{z n(\mathrm{Ze})^{2}}{k_{\mathrm{B}}} \mathrm{e}^{\Delta S_{\mathrm{m}} / k_{\mathrm{B}}} a_{0}^{2} \nu_{0}
\end{equation}
where $z$ is the correlation factor, $n$ is the charge carrier density, Ze is the charge of the mobile ion, $\Delta S_{\mathrm{m}}$ is the entropy of migration, $k_{\mathrm{B}}$ is the Boltzmann constant, $a_{0}$ is the jump distance, and $\nu_0$ is the attempt frequency. Attempt frequency can vary much and influence the prefactor\cite{kraft_influence_2017,li_hopping_2023,krauskopf_comparing_2018}. Klerk et al.\cite{de_klerk_analysis_2018} proposed that the attempt frequency can be approximated by the band center of the mobile cation's phonon density of states (DOS). The calculated phonon DOS and cation band center are shown in Figure~5a. The rotational constraint (absence of anion rotation) softens the \ch{Li^+} sublattice in \ch{LiBH4}, resulting in a decrease in the attempt frequency, which can lead to a decrease in prefactor; the \ch{Li^+} phonon band center exhibits a consistent trend across an extended temperature range, as illustrated in Figure~S15b. Migration entropy also plays a crucial role, being related to the vibrational frequencies of configurations at both equilibrium and the saddle point \cite{vineyard_frequency_1957}. The presence of fast anion rotation can alter the local dynamics of the cation in two configurations, thereby influencing the entropy. This is also consistent with the Meyer-Neldel rule, as the fast rotation of \ch{BH4^-} increases both the barrier accompanied by an increase in prefactor, and the combined effect results in higher conductivity.  These results indicate that each anion motion mode influences the conductivity by dynamically altering both the activation energy and the prefactor, with ion migration being affected through fluctuations in the energy landscape and changes in the local cation dynamics.
\subsection{Anion Dynamics Driving Superionicity via Coupling and Disorder}
We now consider the phonon DOS for each superionic conductor under four constrained conditions (Figure~5a--c and Figure~S16--S21). The phonon DOS calculated from MD trajectories, which naturally accounts for anharmonicity to all orders, can effectively capture the dynamics of the system or be atom-projected to extract cation dynamics. Here, applying constraints to anion dynamics dramatically affects the cation phonon DOS. In different materials, the anion motion mode that dominates the cation phonon DOS also has the most significant effect on conductivity.

For \ch{LiBH4}, when RTVC are applied to \ch{BH4^-}, distinct peaks are observed in the \ch{Li^+} phonon DOS. When only the vibrations of \ch{BH4^-} are present, the \ch{Li^+} phonon DOS is hardly affected, indicating that the high-frequency optical modes decouple from \ch{Li^+} dynamics. However, translation induces a drastic change in the \ch{Li^+} phonon DOS, causing it to become more diffuse and undergo strong intensity suppression (Figure~5a). These indicate a strong coupling between \ch{BH4^-} translation and \ch{Li^+} motion. Anion translation can couple with certain modes of \ch{Li^+}, which are essential for migration, thereby driving superionicity. In addition, damping can also result in a more diffuse phonon DOS due to broadening\cite{maradudin_scattering_1962}; this causes \ch{Li^+} to undergo less well-defined lattice vibrations and exhibit more liquid-like behavior, reminiscent of the breakdown of phonons involving the Cu sublattice in \ch{CuCrSe2} during its superionic transition\cite{ niedziela_selective_2019}. This has also been observed in several other superionic conductors during their superionic transitions \cite{peyrard_temperature_1975, ding_anharmonic_2020}. Meanwhile, anion translation directly triggers the emergence of low-frequency modes of cation motion (see red line in Figure~5a), leading to large atomic displacements and thereby contributing to superionicity.

The influence of anion translation can also be interpreted by mechanism. MD is a good approach for detecting underlying ion transport mechanism. \ch{LiBH4} is characterized by split of Li site along $c$ axis \cite{ikeshoji_diffuse_2009} and a weak anisotropic diffusion pathway in the $ab$ plane and along $c$ axis\cite{aeberhard_ab_2012}. The split site typically serves as an interstitial site through which migration occurs in the ab-plane and along the $c$ axis\cite{ikeshoji_fast-ionic_2011}. Under RC, the split undergoes a minor contraction but persists (Figure~6a). However, under RTC, the doubly split nature is completely suppressed. Additionally, in Figure~6b--c, we compare the \ch{Li^+} probability densities with and without anion translation, clearly illustrating the disappearance of the double splitting when the anion translation is constrained. Another piece of evidence is the pair distribution fucntion (PDF) of \ch{Li^+}--\ch{Li^+} (Figure~S22a), where the SS1--SS2 distribution disappears under RTC. The absence of \ch{BH4^-} translation renders the split site highly energetically unfavorable, thereby blocking the migration pathway.
\begin{figure}[H]
\centering
\includegraphics[width=\textwidth]{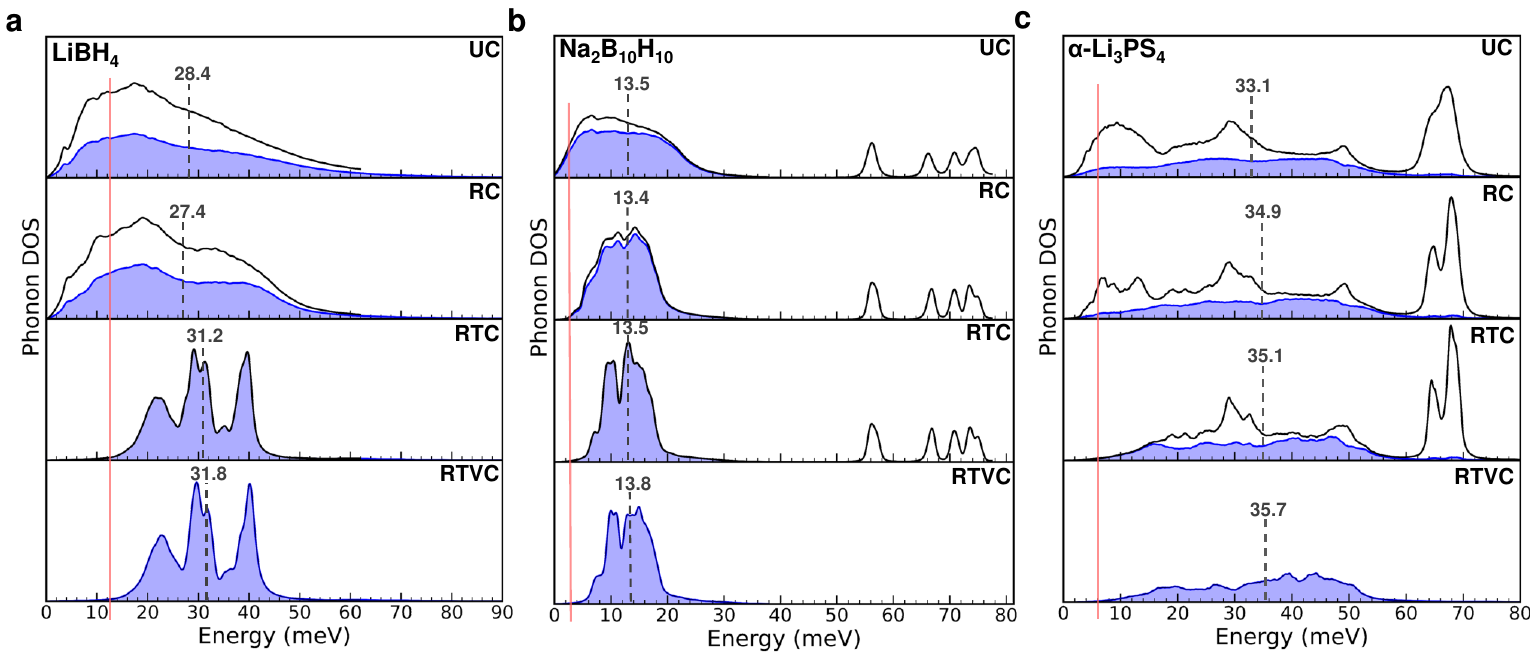}
\caption*{\textbf{Figure\:5.}~Total and cation-projected phonon DOS of (a) \ch{LiBH4} at 550\:K, (b) \ch{Na2B10H10} at 350\:K, and (c) $\alpha$-\ch{Li3PS4} 400\:K under different constraints. The blue shaded areas correspond to the mobile cation-projected phonon DOS. The number and dashed line represent the cation band center. The red line indicates the emergence of the cation phonon DOS low-frequency component when anion dynamics are present. Higher frequency peaks are not plotted for clarity.}
\label{fig5}
\end{figure}
Similar to \ch{LiBH4}, the vibrations of the \ch{B10H10^{2-}} anion have minimal effect on the \ch{Na^+} phonon DOS in \ch{Na2B10H10} (Figure~5b). Anion translation leads to more diffuse \ch{Na^+} phonon DOS. Notably, anion rotation strongly affects phonon modes dominated by sodium ions and triggers the emergence of low-frequency modes of \ch{Na^+} motion. This class of materials is characterized by an order-disorder phase transition (superionic transition), after which anion reorientation becomes facile and partial cation occupancy emerges (Figure~S23). The \ch{Na^+}--\ch{Na^+} PDF (Figure~6d) and \ch{Na^+} probability densities under UC and RC (Figure~6e--f) suggest that anion rotation directly induces cation disorder, leading to the distribution of cations across both tetrahedral and octahedral sites with partial occupancy, thereby opening transport channels. 

For \ch{Li3PS4}, the \ch{Li^+} phonon DOS is diffuse, even within the static anion framework (under RTVC) (Figure~5c). Strikingly, \ch{PS4^{3-}} vibrations induce disorder within \ch{Li^+} sublattice and cause \ch{Li^+} phonon DOS to become diffuse, with the disorder evidenced by the probability densities of Li ions distributing over multiple partial sites (Figure~6h--i) and by the broadening of peaks in the \ch{Li^+}--\ch{Li^+} PDF from RTC to RTVC (Figure~6g). This indicates a strong dynamic coupling between cation dynamics and the lattice's optical modes. Additionally, the gradual broadening of \ch{Li^+}--\ch{Li^+} PDF (Figure~6g) and gradual emergence of partial occupancy (Figure~S24), resulting from anion translation followed by rotation, indicate that these dynamics can also induce \ch{Li^+} disorder. We conclude that \ch{Li^+} motion is strongly coupled with all \ch{PS4^{3-}} motion modes. In two sodium thiophosphates-type SEs, the \ch{Na^+} phonon DOS in \ch{Na4P2S6} display an active response to \ch{P2S6^{4-}} vibrations (Figure~S20), whereas in \ch{Na11Sn2PS12}, the \ch{Na^+} phonon DOS are less sensitive to \ch{PS4^{3-}} dynamics (Figure~S21).

Across the entire simulated temperature range, we find that anion dynamics lower the cation band center value, effectively softening the cation sublattice (Figure~S15). The cation band center was correlated with migration enthalpy\cite{muy_tuning_2018} and utilized as a descriptor to identify promising ionic conductors\cite{muy_high-throughput_2019}. Here, the anion dynamics facilitate a softer cation sublattice and thus lower migration enthalpy. We calculate the difference in cation band center value between UC and RTVC, as shown in Figure~S25. However, \ch{B10H10^{2-}} dynamics have a much smaller softening effect on the \ch{Na^+} sublattice in \ch{Na2B10H10}, yet dominate ion migration (Figure~3a,f). On one hand, the cation band center represents an averaged lattice vibration property that characterizes the lattice softness, i.e., the interactions strength between atoms. On the other hand, certain anion motion modes, though not significantly softening the cation sublattice, couple with cation motions that are essential for migration. Identifying these anion dynamics, as the constraint algorithm enables us to do here, and selectively exciting them provide fundamental insights and enable high ion mobility. 
\begin{figure}[H]
\centering
\includegraphics[scale=.73]{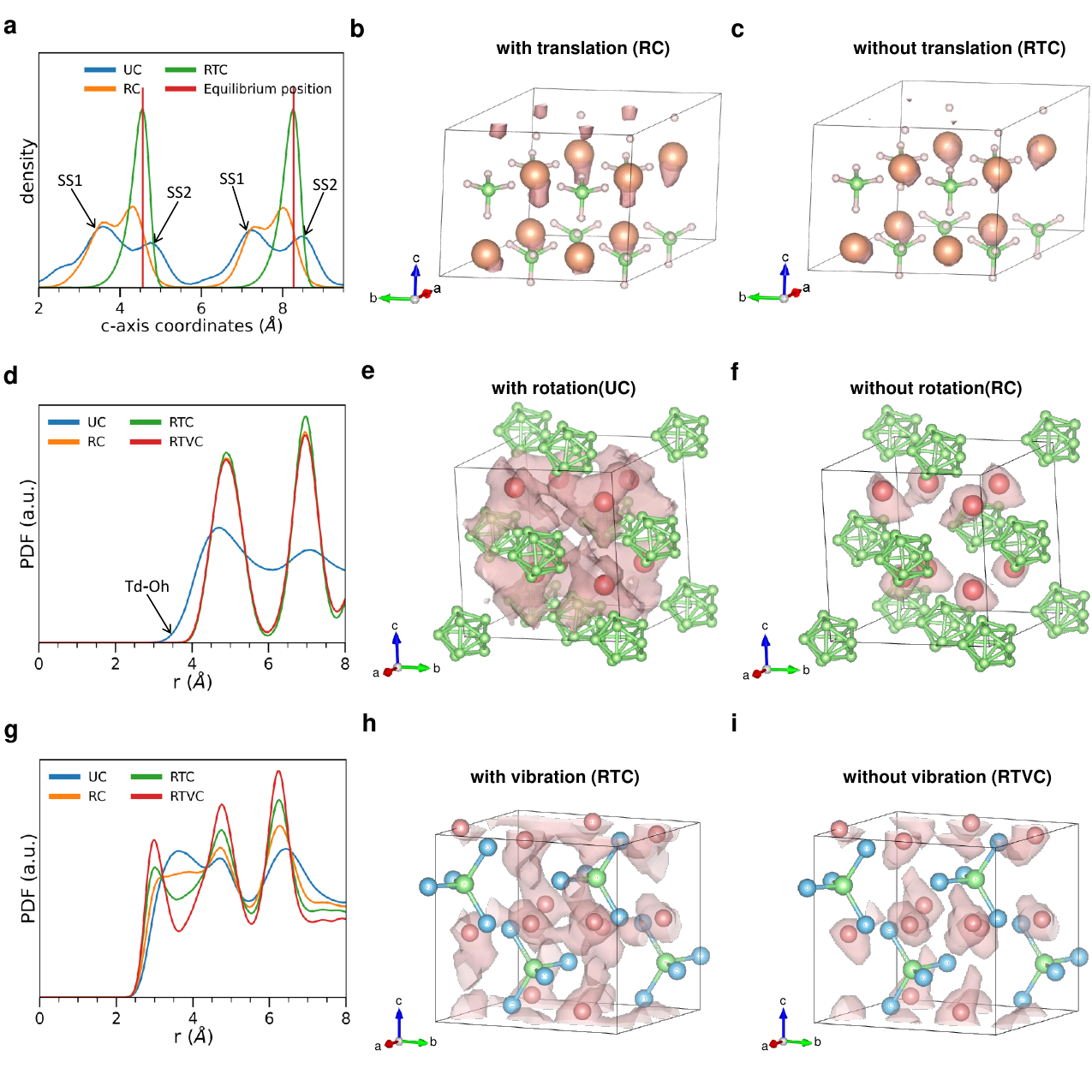}
\caption*{\textbf{Figure\:6.}~(a) Statistical distribution of the Li ions along $c$ axis in \ch{LiBH4} under UC, RC, and RTC at 550\:K, along with the equilibrium lattice positions. The distribution under RTVC overlapping with that under RTC is not shown. The probability densities of Li ions in \ch{LiBH4} obtained from MLMD simulations at 550\:K: (b) with \ch{BH4^-} translation (RC) and (c) without \ch{BH4^-} translation (RTC), shown with an isosurface value of $0.045$. (d) PDF of Na ions in \ch{Na2B10H10} under four constrained conditions, calculated from MLMD trajectories at 650\:K. The probability densities of Na ions in \ch{Na2B10H10} at 750\:K: (e) with \ch{B10H10^{2-}} rotation (UC) and (f) without \ch{B10H10^{2-}} rotation (RC), shown with an isosurface value of $7.0 \times 10^{-4}$; the experimentally determined partial occupancies of tetrahedral (Td) and octahedral sites (Oh) are 0.58 and 0.28, respectively.\cite{udovic_exceptional_2014}. (g) PDF of Li ions in \ch{Li3PS4} under RTC and RTVC, calculated from MLMD trajectories at 550\:K. The probability densities of Li ions in \ch{Li3PS4} at 600\:K: (h) with \ch{PS4^{3-}} vibration (RTC) and (i) without \ch{PS4^{3-}} vibration (RTVC), shown with an isosurface value of $0.001$.}
\label{fig6}
\end{figure}
\section{Conclusions}
We trained MLIPs for six plastic phases across three distinct classes of materials and successfully reproduced cation diffusion and facile anion rotation at slow, moderate, and fast frequencies. Ultralong-time MLMD simulations were conducted for \ch{Na2B10H10}, \ch{LiBH4}, \ch{Na3OBH4}, $\alpha$-\ch{Li3PS4}, $\gamma$-\ch{Na4P2S6}, and \ch{Na11Sn2PS12} at near room temperature and elevated temperature. Using our newly developed constraint algorithm, we quantitatively evaluated the impact of each anion motion mode—rotation, translation, and vibration—on conductivity. We found that anion dynamics are of paramount importance for enabling superionicity, especailly at lower temperatures, both by inducing fluctuations in the underlying energy landscape, thereby effectively lowering the energy barrier, and by altering local cation dynamics, which directly affects the prefactor. 

More importantly, we find that different anion motion modes dominate in various superionic conductors. In \ch{LiBH4}, although the anion undergoes quasi-free rotation, cation diffusion is highly coupled with the translational degrees of freedom of the \ch{BH4^-} anion. The absence of translation causes this material to behave as an `insulator' within the simulation timescale. In \ch{Na2B10H10}, the hopping frequency of sodium ions matches the reorientation frequency of the semispherical anion. This moderate-frequency reorientation primarily governs cation diffusion in this material. In contrast, in $\alpha$-\ch{Li3PS4}, anion vibration plays a significant role in cation diffusion. These dominant anion motion modes influence cation motion by inducing disorder within the cation sublattice, and their effect on ion mobility becomes more prominent at lower temperatures. This study provides a fundamental understanding of anion-cation coupling dynamics and how subtle tuning of polyanion dynamics can be the key to achieving high conductivity, particularly at lower temperatures. We hope that these new insights will pave the way for exploring future superionic conductors.

\bibliography{achemso-demo}

\end{document}

%\begin{tocentry}
%Some journals require a graphical entry for the Table of Contents.
%This should be laid out ``print ready'' so that the sizing of the

%Inside the \texttt{tocentry} environment, the font used is Helvetica
%8\,pt, as required by \emph{Journal of the American Chemical

%The surrounding frame is 9\,cm by 3.5\,cm, which is the maximum
%permitted for  \emph{Journal of the American Chemical Society}
%graphical table of content entries. The box will not resize if the
%content is too big: instead it will overflow the edge of the box.

%This box and the associated title will always be printed on a
%separate page at the end of the document.
%%\captionsetup[subfloat]{font={bf,small}, skip=1pt, margin=-0.7cm, singlelinecheck=false}
%\end{tocentry}

\newpage

\subsection{1.\:Machine Learning Interatomic Potential}
Geometry optimizations and single-point calculations were performed using the projector augmented-wave (PAW)\cite{blochl_projector_1994,perdew_generalized_1996} method and the Perdew-Burke-Ernzerhof (PBE)\cite{perdew_generalized_1996} functional in the Vienna ab initio simulation package (VASP) 5.4.4\cite{kresse_ultrasoft_1999,kresse_efficient_1996}. The convergence tests for K-point sampling and energy cutoff were carried out to ensure that forces are converged to less than 1 meV/\AA. Due to the relatively small number of atoms in the unit cells of \ch{LiBH4} and \ch{Na3OBH4}, we begin with their 2$\times$2$\times$1 and 2$\times$2$\times$2 supercells, respectively. Structure manipulation was dealt with \textit{pymatgen}\cite{ong_python_2013}. 

The DP-GEN\cite{zhang_dp-gen_2020} scheme was adopted for the deep potential (DP) \cite{zeng_deepmd-kit_2023} model generation. For each material, a dataset containing 6,000 DFT energies and forces was used to train the initial model. During each iteration, four models are generated, and the \(\sigma_{high}\) and \(\sigma_{low}\) are set to 0.26 and 0.11, respectively, to evaluate model deviation. A cut-off radii of 7Å is selected for training, with 400,000 training steps. The exploration phase uses LAMMPS software\cite{plimpton_fast_1995}, with MD steps increasing from 1500 to 10,000, after which hundreds of candidate configurations are chosen for labeling. The exploration of each system is considered converged when the percentage of accurate configurations reaches 99.5\%. After the models trained on unit cells (supercells for \ch{LiBH4} and \ch{Na3OBH4}) converged, we expanded \ch{Na2B10H10}, \ch{LiBH4}, \ch{Na3OBH4}, \ch{Li3PS4}, and \ch{Na4P2S6} to 2$\times$2$\times$1, 3$\times$3$\times$2, 3$\times$3$\times$3, 2$\times$2$\times$2, and 2$\times$2$\times$2 supercells, respectively, repeating the process until the same convergence criteria were met. For \ch{Na11Sn2PS12}, which already has a large number of atoms in its unit cell and is highly time-consuming, no expansion was performed. Finally, six model ensembles, each containing four models, were obtained.

Bofore utilizing MLIPs for predicting ion migration, thorough and careful testing is essential. In general, a good model should reproduce key observations and yield low numerical error. In this case, the rotation of four distinct rotors across six materials was observed, showing frequencies similar to those simulated in first-principle calculations and/or measured in the experiments.\cite{remhof_rotational_2010,zhang_coupled_2019,scholz_superionic_2022,udovic_exceptional_2014,dimitrievska_carbon_2018,sun_rotational_2019,buchter_dynamical_2008,zhang_targeting_2020} %(see Figure~2 in next section) %To further assess the model performance, we calculate the root mean square error (RMSE) of energies and forces. The results are listed in Table~\ref{TableDP}. 
As suggested by Morrow et al.,\cite{morrow_how_2023} a smaller system size results in a higher energy RMSE per atom. To avoid underestimating error while saving computational costs, we select a smaller system size than that used in training for generating the test sets.
%The RMSE for energies and forces is around 1 meV per atom and several tens of meV/\AA\:respectively, showing a significant improvement over the empirical interatomic potentials. The low standard deviations in both energies and forces indicate the uniform accuracy of the four models. Meanwhile, a small disparity between test and training data suggests no explicit overfitting. 
Furthermore, it is important to ensure that the model performs reasonably well across as much of the configuration space as possible. Therefore, diagnostic plots comparing the DFT and DP models are shown in Figure~S1--S2 to visualize the error distribution. These plots demonstrate that our MLIPs are reliable and accurately reproduce the DFT energies and forces over a wide range of configurational space.

\subsection{2.\:Molecular Dynamics Simulation}
MD simulations for production of diffusivity with high accuracy over a wide range of temperatures and all constrained scenarios were performed with CP2K package\cite{kuhne_cp2k_2020}. The systems were first equilibrated for 10\:ps, after which a\:10 ns simualtion within the canonical (NVT) ensemble using a Nose thermostat\cite{nose_molecular_1984,nose_unified_1984} were carried out. Volumes were fixed to zero-temperature atomic
 relaxations unless stated otherwise.  The detailed molecular dynamics settings for each system were listed in Table~S1.
\begin{table}[]
  \caption*{\textbf{Table\:S1.}~MD settings for six electrolytes}
\label{MDSettings}
\begin{tabular}{llll }
\toprule
Materials   &Size ($N_{atoms}$)  & Timestep (fs) & Temperatures (K)\\
\midrule
\ch{Na2B10H10} & \(3 \times 3 \times 3\) (1056) & 0.5 & 750, 700, 650, 550, 450, 350\\
\ch{LiBH4} & \(4 \times 4 \times 4\) (768) & 0.5 & 600, 550, 525, 500, 475, 450, 350\\
\ch{Na3OBH4} & \(5 \times 5 \times 5\) (1125) & 0.5 & 800, 550, 450\\
\ch{Li3PS4} & \(3 \times 3 \times 3\) (864) & 2.0 & 1100, 1000, 900, 800, 700, 600, 550, 500, 400\\
\ch{Na4P2S6} & \(3 \times 3 \times 3\) (648) & 2.0 & 1000, 950, 900, 850, 800, 750 \\
\ch{Na11Sn2PS12} & \(2 \times 2 \times 1\) (832) & 2.0 & 1200, 1050, 900, 750, 600, 500, 400, 300 \\
\bottomrule
\end{tabular}
\end{table}

The conductivity is calculated by Nernst-Einstein relation: 
\begin{equation}
\sigma = \frac{Nq^{2}}{Vk_{B}T}D_{tr}
\label{cond}
\end{equation}
where \(N\) and \(q\) are the number and electric charge of transport ions, respectively. \(V\) is the volume of the simulation box, \(k_{B}\) is the Boltzmann constant, and \(T\) is the temperature. The tracer diffusion coefficient \(D_{tr}\) is defined as
\begin{equation}
D_{t r}=\lim _{t \rightarrow \infty} \frac{\frac{1}{N} \sum_{i}^{N}<\left[\boldsymbol{r}_{\boldsymbol{i}}(t)-\boldsymbol{r}_{\boldsymbol{i}}(0)\right]^{2}>}{6 t}
\end{equation}
where \(\boldsymbol{r}_{i}(t)\) denotes the position of \textit{i}th mobile ion at time t and 6 comes from the 3D nature of the system. First take the average over different initial time, then take the average over N charge carrier, which gives us average mean square displacement. \(D_{tr}\) is then readily obtained by a linear fitting of \(t\) and MSD. 
When the diffusion coefficient is around $10^{-7}$\:cm$^{2}$\:s$^{-1}$, MLMD simulations of thousands of atoms lasting 10\:ns typically yield results with minimal error\cite{huang_deep_2021}. Here, we choose $10^{-8}$\:cm$^{2}$\:s$^{-1}$ as the threshold, below which the obtained diffusion coefficient is considered invalid. $10^{-8}$\:cm$^{2}$\:s$^{-1}$ typically corresponds to a conductivity on the order of 0.1\:mS\:cm$^{-1}$.

The percentage ratio of conductivity reduction caused by constraining each anion motion mode is given by
\begin{equation}
r_{mode}=\frac{\Delta_{mode}}{\Delta_{Rot}+\Delta_{Trs}+\Delta_{Vib}}
%\quad \text{where } mode \in \{Rot, Trs, Vib\}
\end{equation}
in which $\Delta_{mode}$ is the conductivity reduction caused by constraining a specific mode and can be one of $\Delta_{Rot}\text{, }\Delta_{Trs}\text{, or }\Delta_{Vib}$, each calculated by
\begin{align}
&\Delta_{Rot}=\log(\sigma_{UC}) - \log(\sigma_{RC}) \\
&\Delta_{Trs}=\log(\sigma_{RC}) - \log(\sigma_{RTC})\\
&\Delta_{Vib}=\log(\sigma_{RTC}) - \log(\sigma_{RTVC})
\end{align}
in which the $\sigma_{UC}$, $\sigma_{RC}$, $\sigma_{RTC}$, and $\sigma_{RTVC}$ are conductivity estimated using eq~\ref{cond} under different constrained cases. %For the estimation of the contribution of \ch{B10H10^{2-}} rotation to conductivity at 550\:K, the diffusion coefficient is typically below $10^{-8}$\:cm$^{2}$\:s$^{-1}$. We use 0.01\:mS\:cm$^{-1}$ as the lower limit for this estimation. In cases where calculated tracer diffusion coefficient is below $10^{-8}$\:cm$^{2}$\:s$^{-1}$, which typically corresponds to a conductivity on the order of $10^{-2}$\:mS\:cm$^{-1}$ , calculating conductivities is not feasible; we use $10^{-2}$\:mS\:cm$^{-1}$ as the reference conductivity value to estimate the percentage ratio of conductivity reduction results from constraining a specific anion motion mode.

The phonon density of states is calculated by\cite{brehm_travisfree_2020}
\begin{equation}
\operatorname{PDOS}(\omega)=\int \frac{\left\langle\sum_{s} v_{s}\left(t_{0}\right) \cdot v_{s}\left(t_{0}+t\right)\right\rangle}{\left\langle\sum_{s} v_{s}\left(t_{0}\right) \cdot v_{s}\left(t_{0})\right\rangle\right.} e^{- i \omega t} dt
\end{equation}
where $s$ runs over all atoms. If it runs over a specific type of atom, the atom-projected PDOS is obtained.

The band center is defined as 
\begin{equation}
\omega_{av}=\frac{\int{\omega}\times PDOS(\omega)d\omega }{\int{PDOS(\omega)d\omega }}
\end{equation}

%The percentage ratio of linewidth reduction caused by constraining each anion motion mode is given by
%\begin{equation}
%r_{mode}=\frac{\Gamma_{mode}}{\Gamma_{Trs}+\Gamma_{Rot}+\Gamma_{Vib}}
%\end{equation}
%where $\Gamma_{mode}$ is the linewidth reduction caused by constraining a specific mode. We calculate the $\Gamma_{Vib}$ by subtracting the linewidth of cation phonon DOS under RTVC from RTC, the $\Gamma_{Trs}$ by subtracting RTC from RC, and the $\Gamma_{Rot}$ by subtracting RC from UC. 
\begin{figure}[H]
\centering
\includegraphics[width=\textwidth]{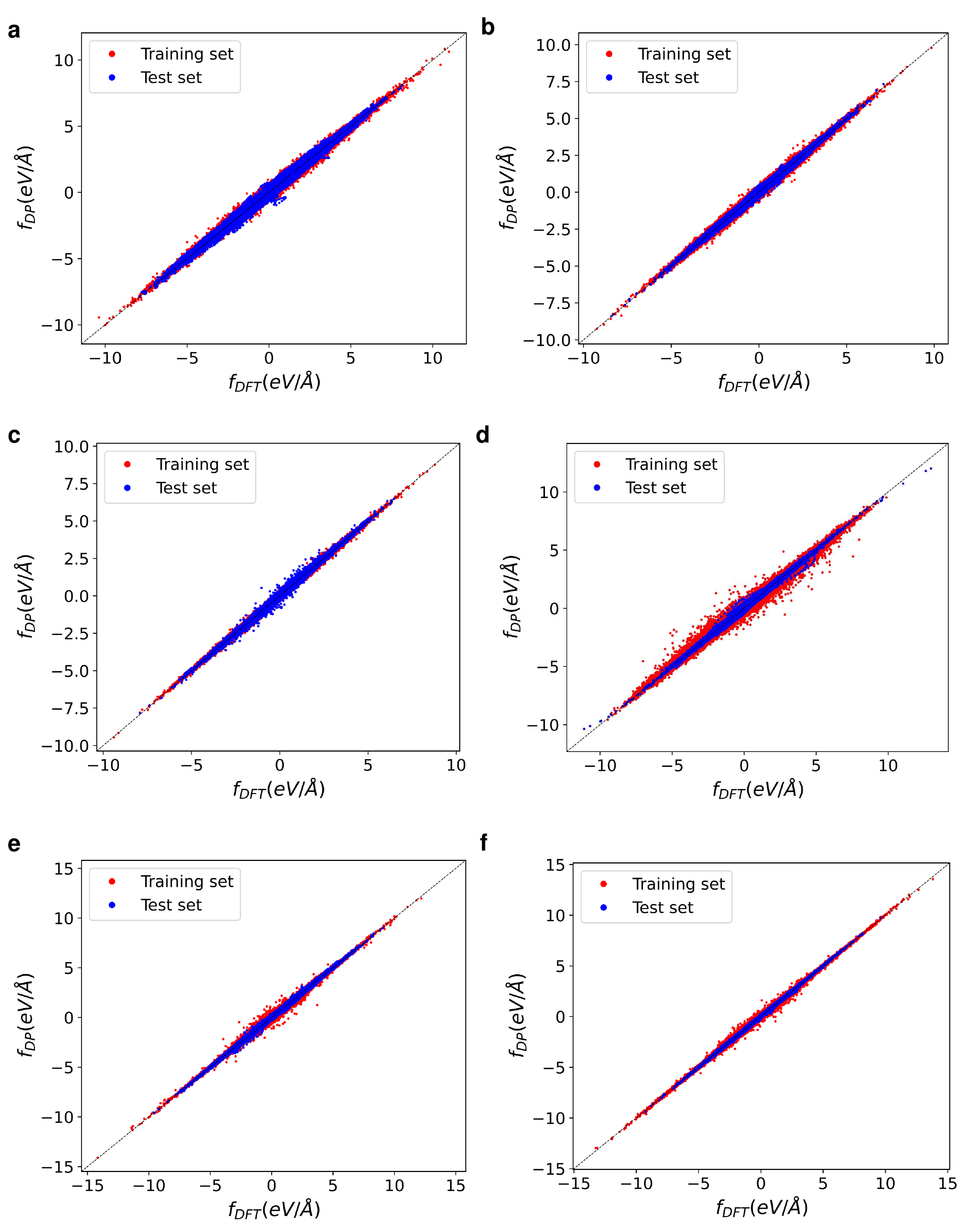}
\caption*{\textbf{Figure\:S1.}~Comparision between DP and DFT forces for (a) \ch{Li3PS4}, (b) \ch{Na4P2S6}, (c) \ch{Na11Sn2PS12}, (d) \ch{Na2B10H10}, (e) \ch{LiBH4}, and (f) \ch{Na3OBH4}.}
\label{figS1}
\end{figure}

\begin{figure}[H]
\centering
\includegraphics[width=\textwidth]{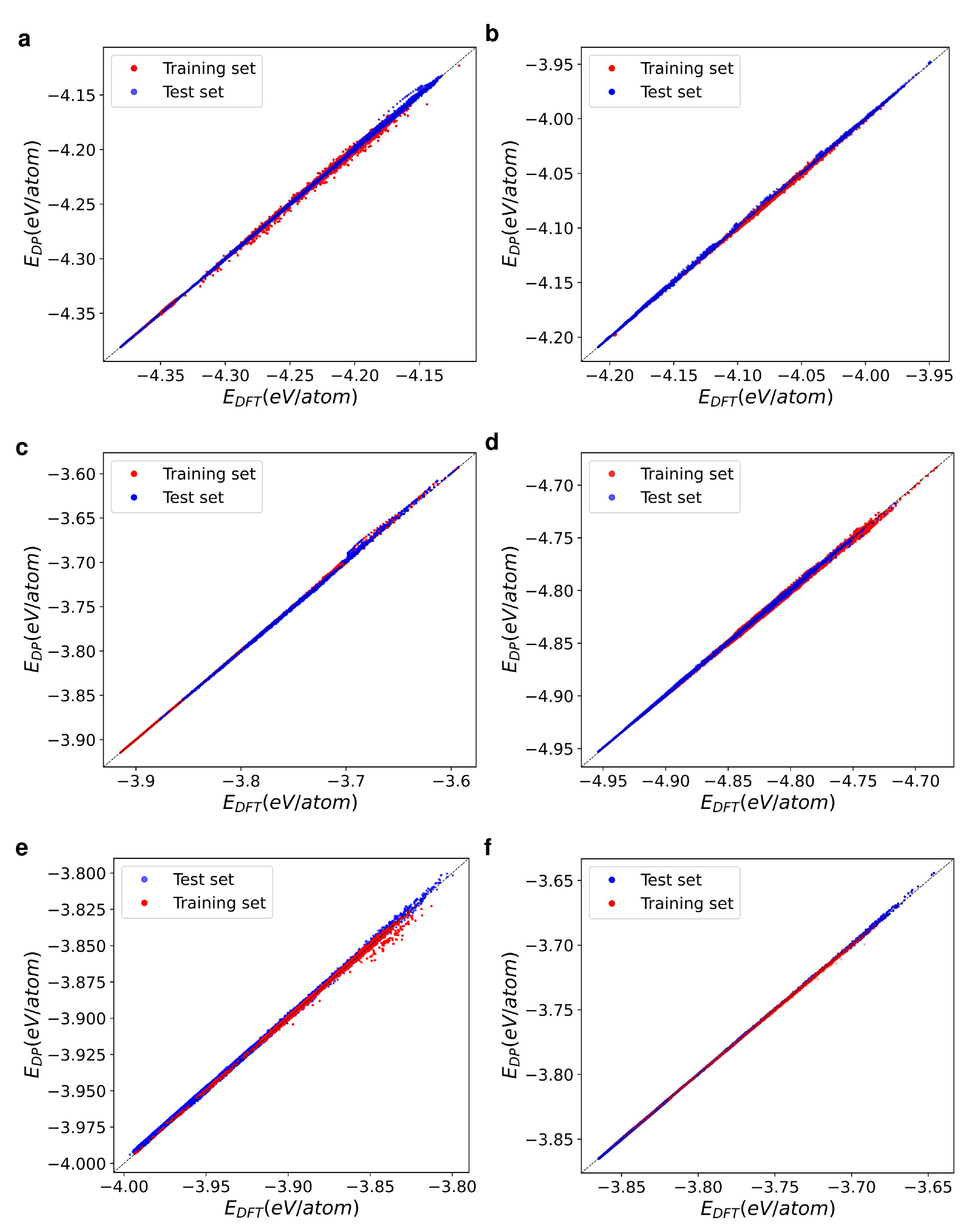}
\caption*{\textbf{Figure\:S2.}~Comparision between DP and DFT energies for (a) \ch{Li3PS4}, (b) \ch{Na4P2S6}, (c)\ch{Na11Sn2PS12}, (d)\ch{Na2B10H10}, (e) \ch{LiBH4}, and (f) \ch{Na3OBH4}.}
\label{figS2}
\end{figure}

\begin{figure}[H]
\centering
\includegraphics[width=\textwidth]{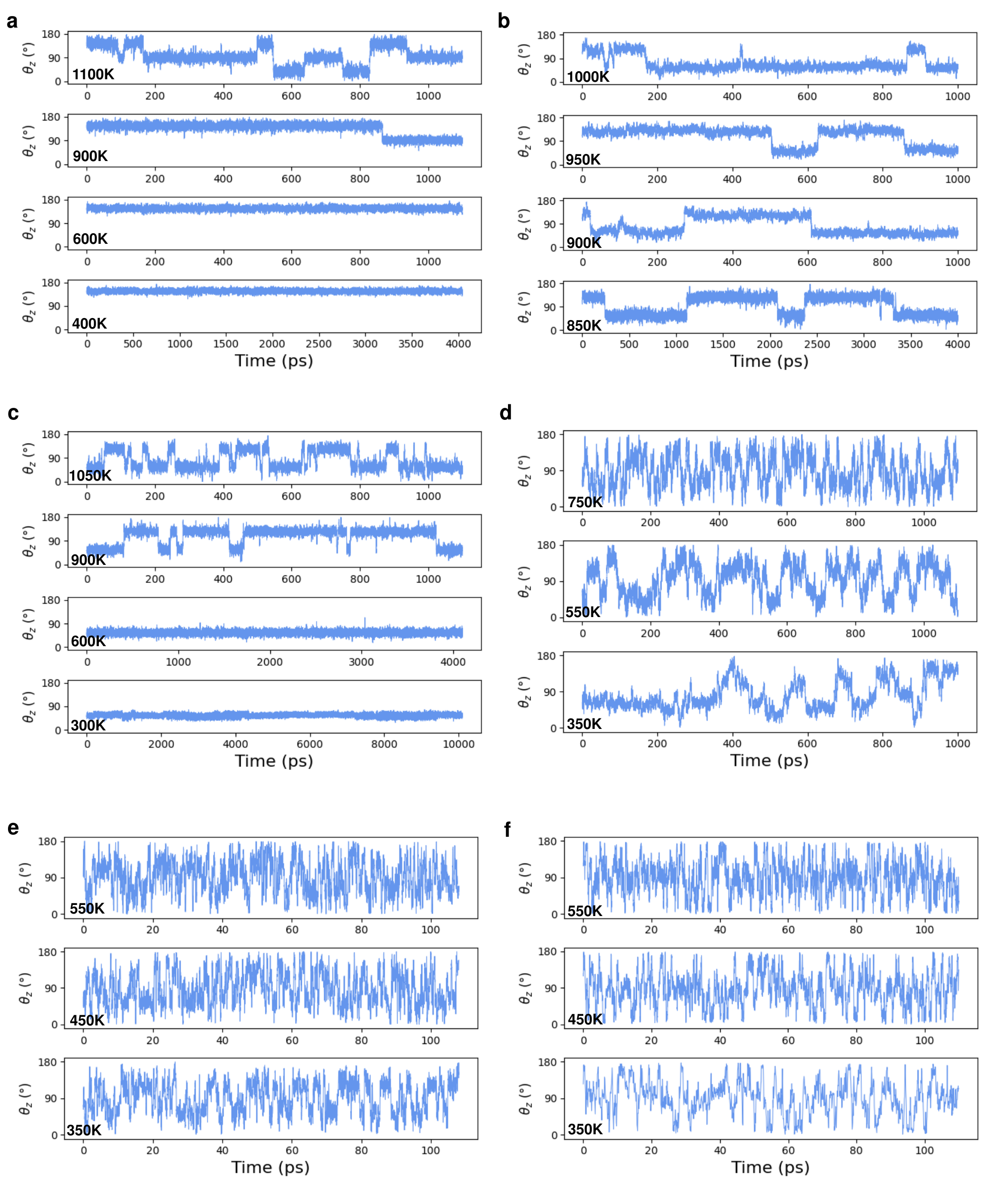}
\caption*{\textbf{Figure\:S3.}~The \(\theta_\alpha\) angle projection for rotor anions in (a) \ch{Li3PS4}, (b) \ch{Na4P2S6}, (c) \ch{Na11Sn2PS12}, (d) \ch{Na2B10H10}, (e) \ch{LiBH4}, and (f) \ch{Na3OBH4} at different temperatures. The angle fluctuations increase with temperature.}
\label{figS3}
\end{figure}

\begin{figure}[H]
\centering
\includegraphics[width=\textwidth]{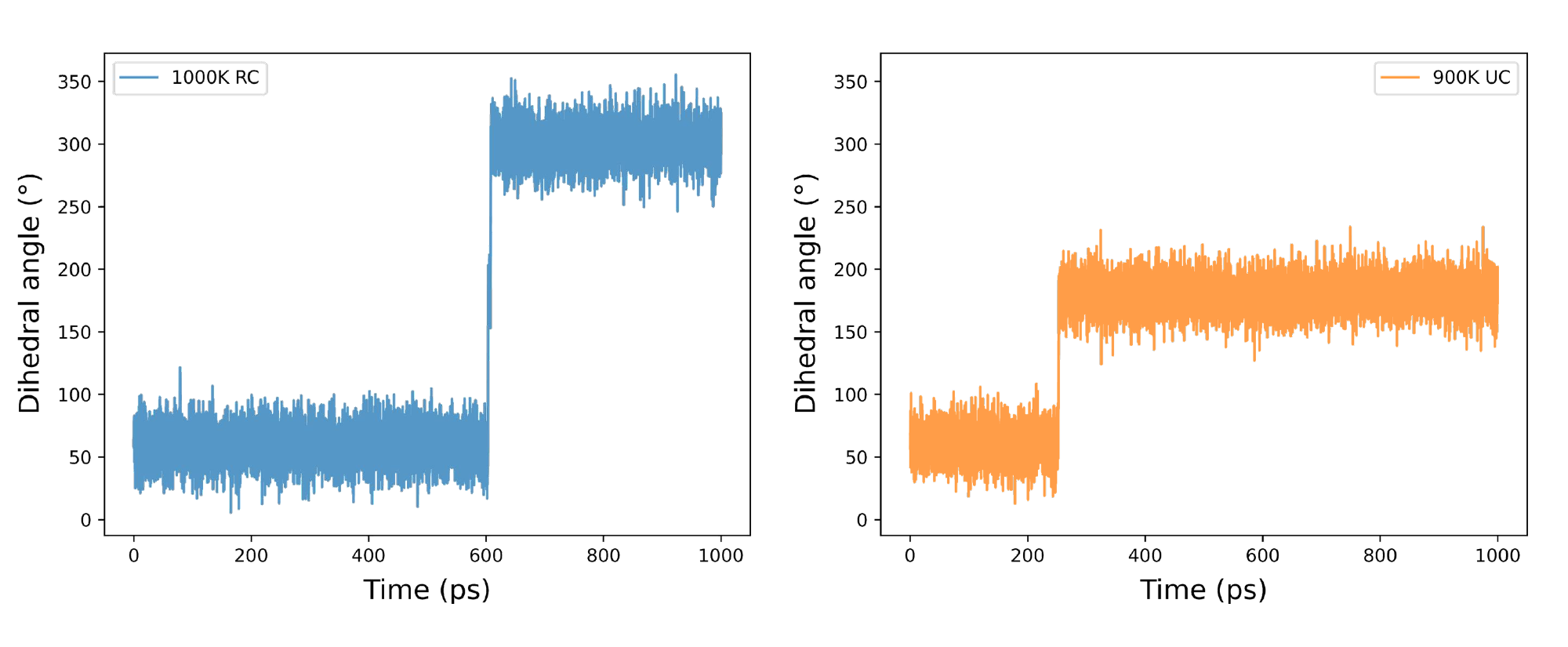}
\caption*{\textbf{Figure\:S4.}~Detected dihedral angles jump of \ch{P2S6^{4-}} in \ch{Na4P2S6} (a) under RC at 1000 K and (b) under UC at 900 K.}
\label{figS8}
\end{figure}

\begin{figure}[H]
\centering
\includegraphics[width=\textwidth]{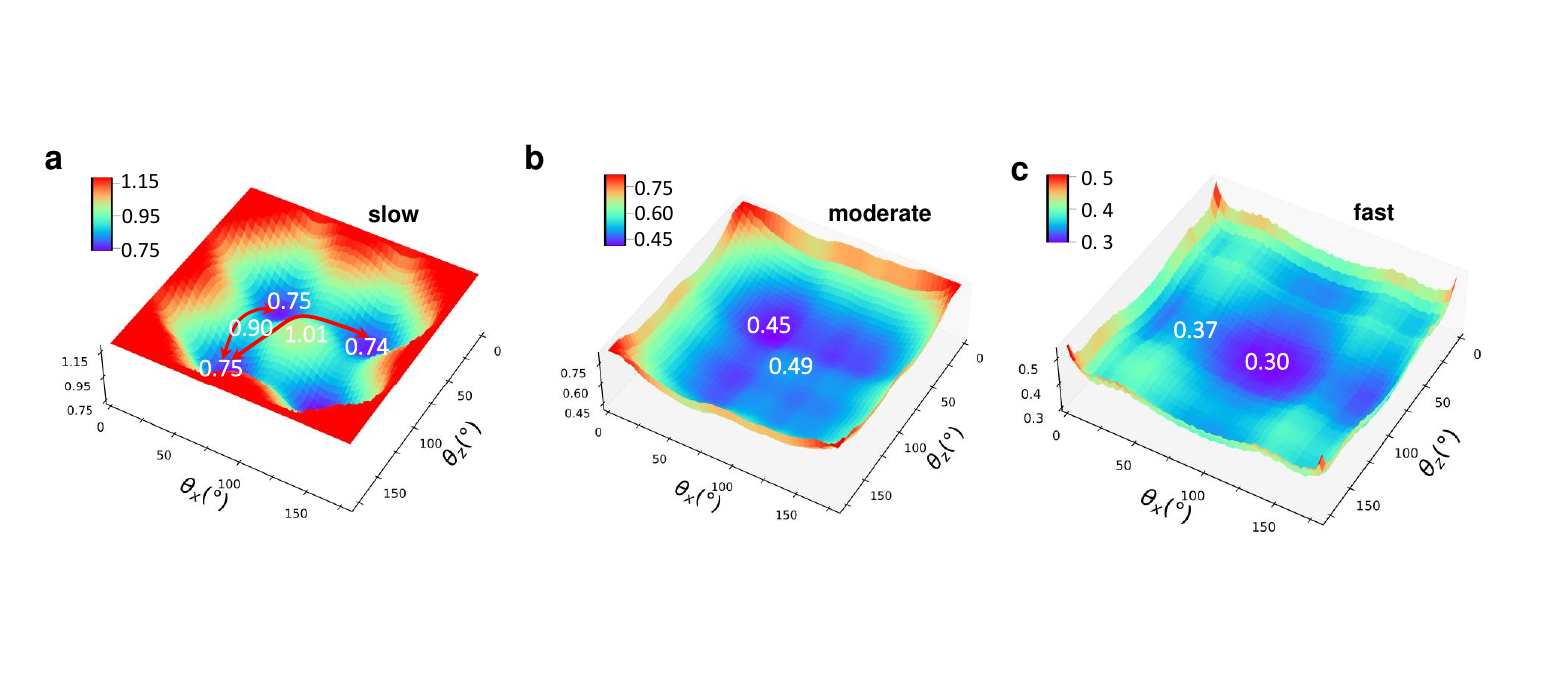}
\caption*{\textbf{Figure\:S5.}~Free energy surfaces for anion rotation of (a) the slow rotor \ch{PS4^{3-}} in \ch{Na11Sn2PS12} at 1200 K, (b) the moderate rotor \ch{B10H10^{2-}} in \ch{Na2B10H10} at 650 K, and (c) the fast rotor \ch{BH4^-} in \ch{Na3OBH4} at 450 K.}
\label{figS4}
\end{figure}
\newpage
\begin{figure}[H]
\centering
\includegraphics[width=\textwidth]{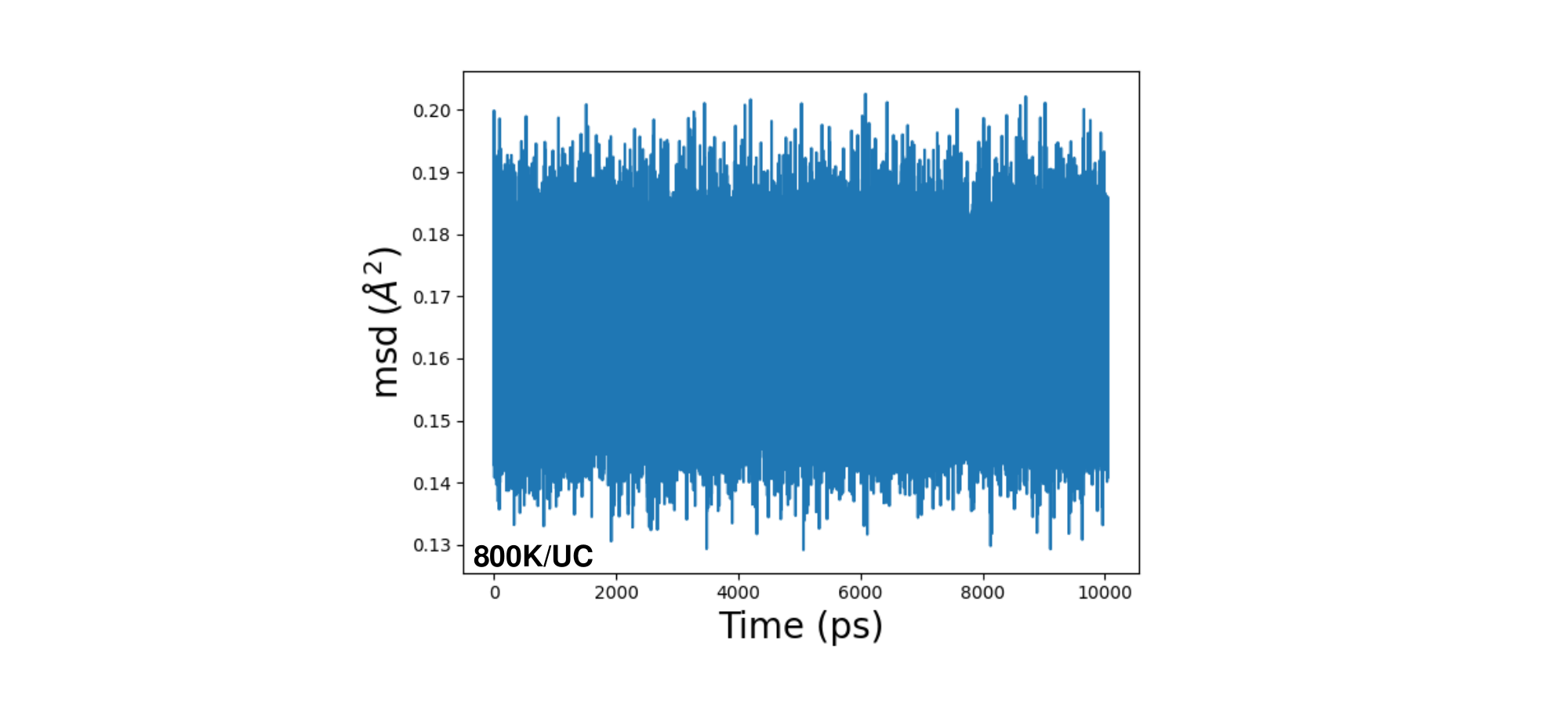}
\caption*{\textbf{Figure\:S6.}~Mean square displacement (MSD) of the \ch{Na^+} in \ch{Na3OBH4} at 800 K with no constraint (unconstraint).}%
\label{figS5}
\end{figure}

\begin{figure}[H]
\centering
\includegraphics[width=\textwidth]{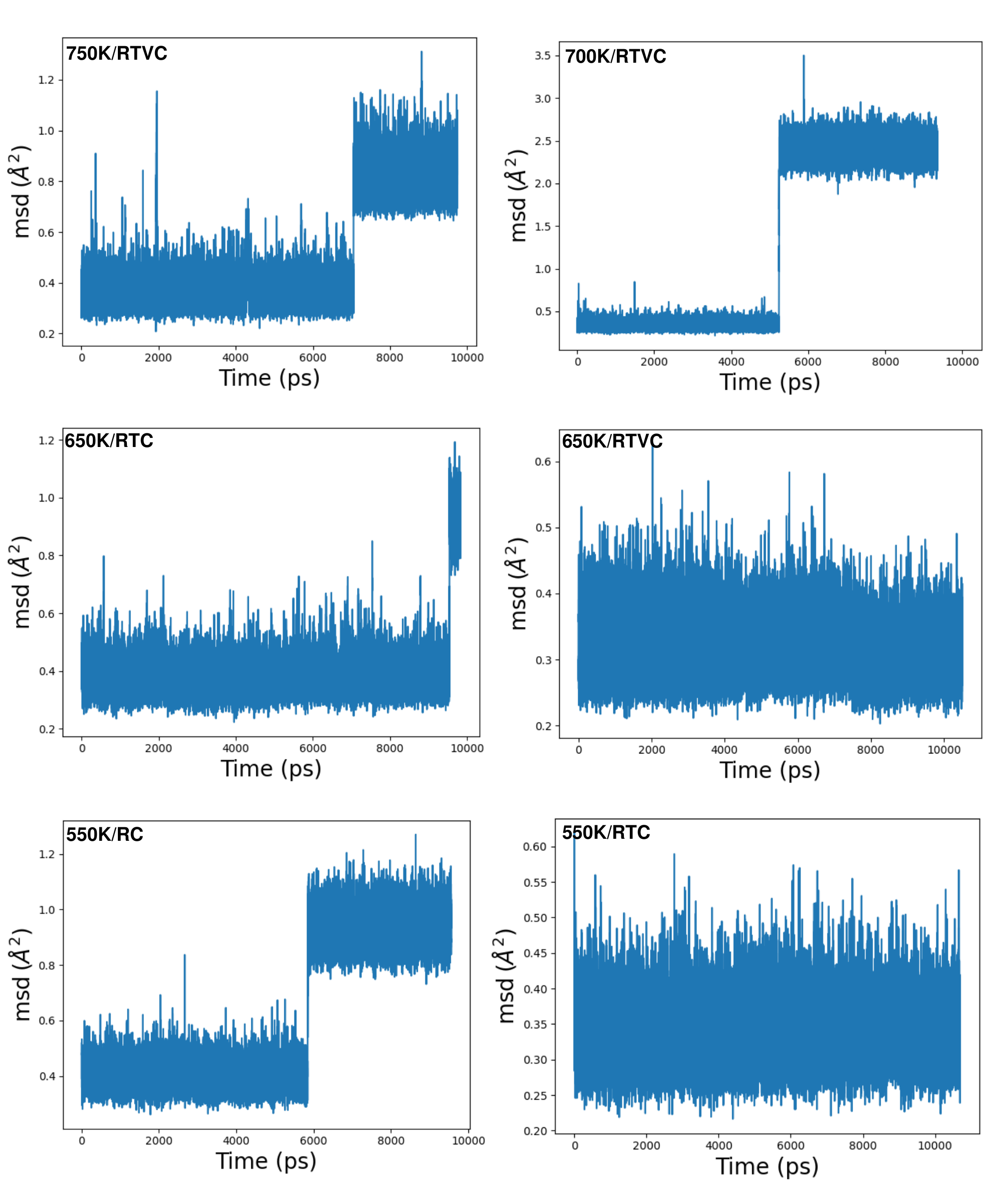}
\caption*{\textbf{Figure\:S7.}~MSD of the \ch{Na^+} in \ch{Na2B10H10} at various temperatures and under different constraints.}%
\label{figS5}
\end{figure}

\begin{figure}[H]
\centering
\includegraphics[width=\textwidth]{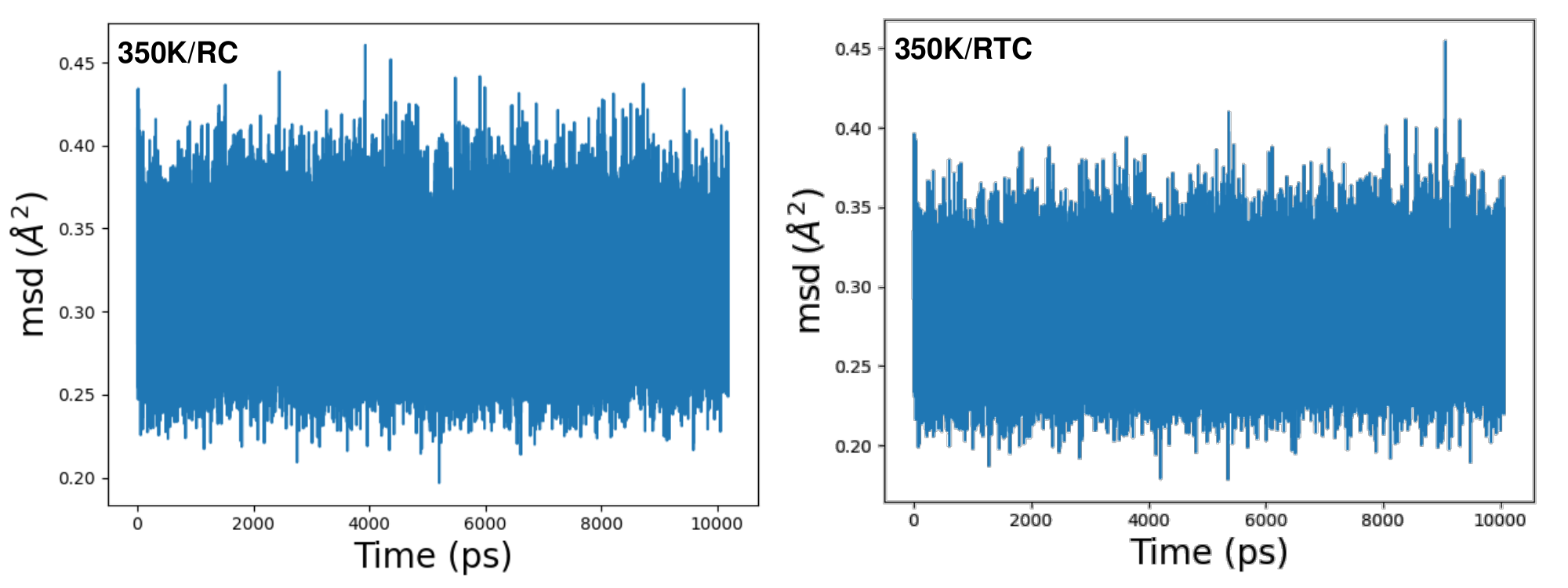}
\caption*{\textbf{Figure\:S8.}~MSD of the \ch{Na^+} in \ch{Na2B10H10} at various temperatures and under different constraints.}%
\label{figS6}
\end{figure}

\begin{figure}[H]
\centering
\includegraphics[width=\textwidth]{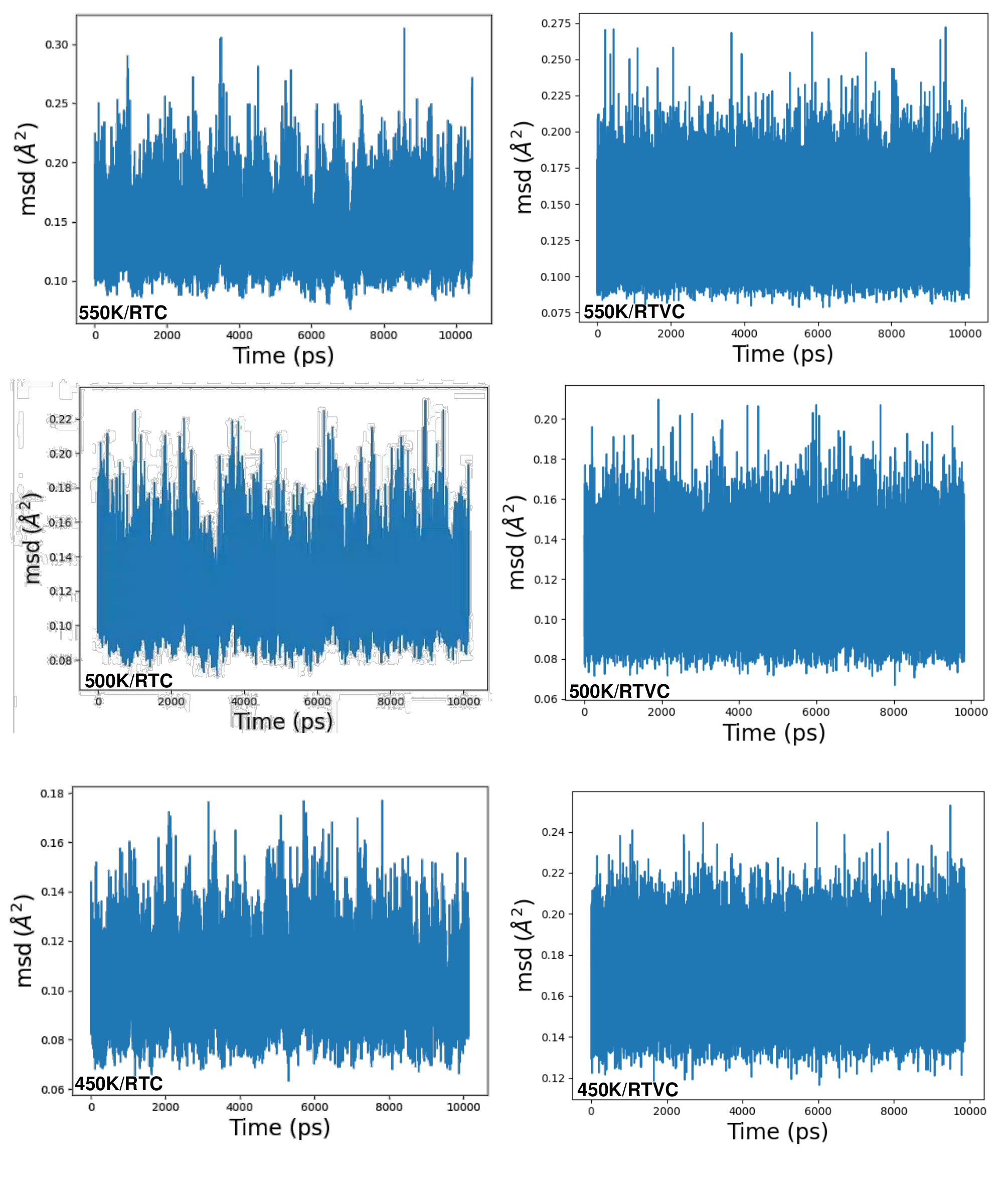}
\caption*{\textbf{Figure\:S9.}~MSD of the \ch{Li^+} in \ch{LiBH4} at various temperatures and under different constraints.}%
\label{figS6}
\end{figure}

\begin{figure}[H]
\centering
\includegraphics[width=\textwidth]{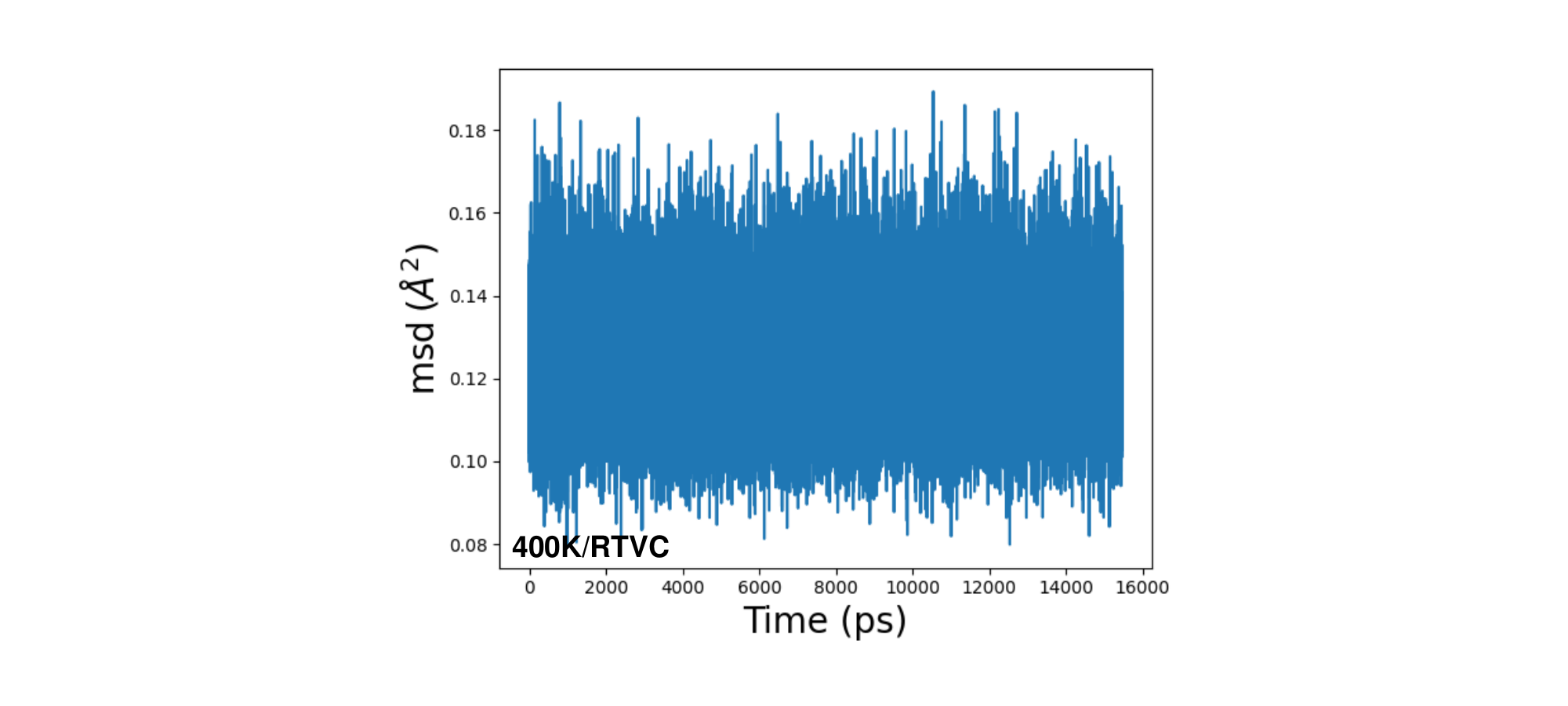}
\caption*{\textbf{Figure\:S10.}~MSD of the \ch{Li^+} in \ch{Li3PS4} at 400\:K under RTVC.}%
\label{figS6}
\end{figure}

\newpage

\begin{figure}[H]
\centering
\includegraphics[width=\textwidth]{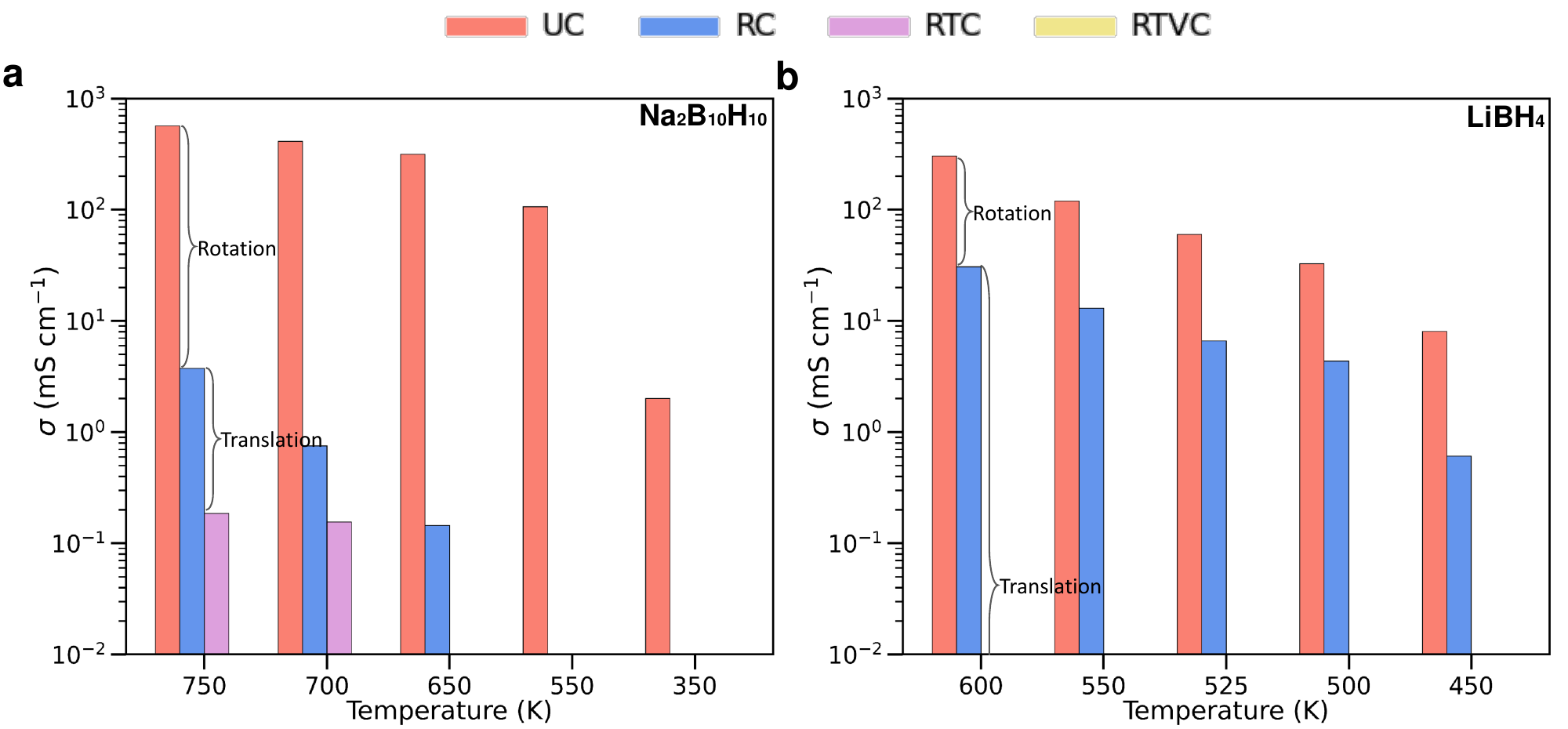}
\caption*{\textbf{Figure\:S11.}~Conductivity of (a) \ch{Na2B10H10} and (b) \ch{LiBH4} in logarithmic scale. For \ch{Na2B10H10} and \ch{LiBH4}, the anion rotation and translation remain the dominant factor influencing conductivity across simulated temperatures, respectively. Translation of \ch{BH4^-} dominates \ch{Li^+} diffusion, and constraining it will cause conductivity to decrease below the measurable range. The percentage contributions of the anion's translation, rotation, and vibration to conductivity cannot be determined because their conductivities under RTVC cannot be computed.}
\label{figS2}
\end{figure}

\begin{figure}[H]
\centering
\includegraphics[scale=.65]{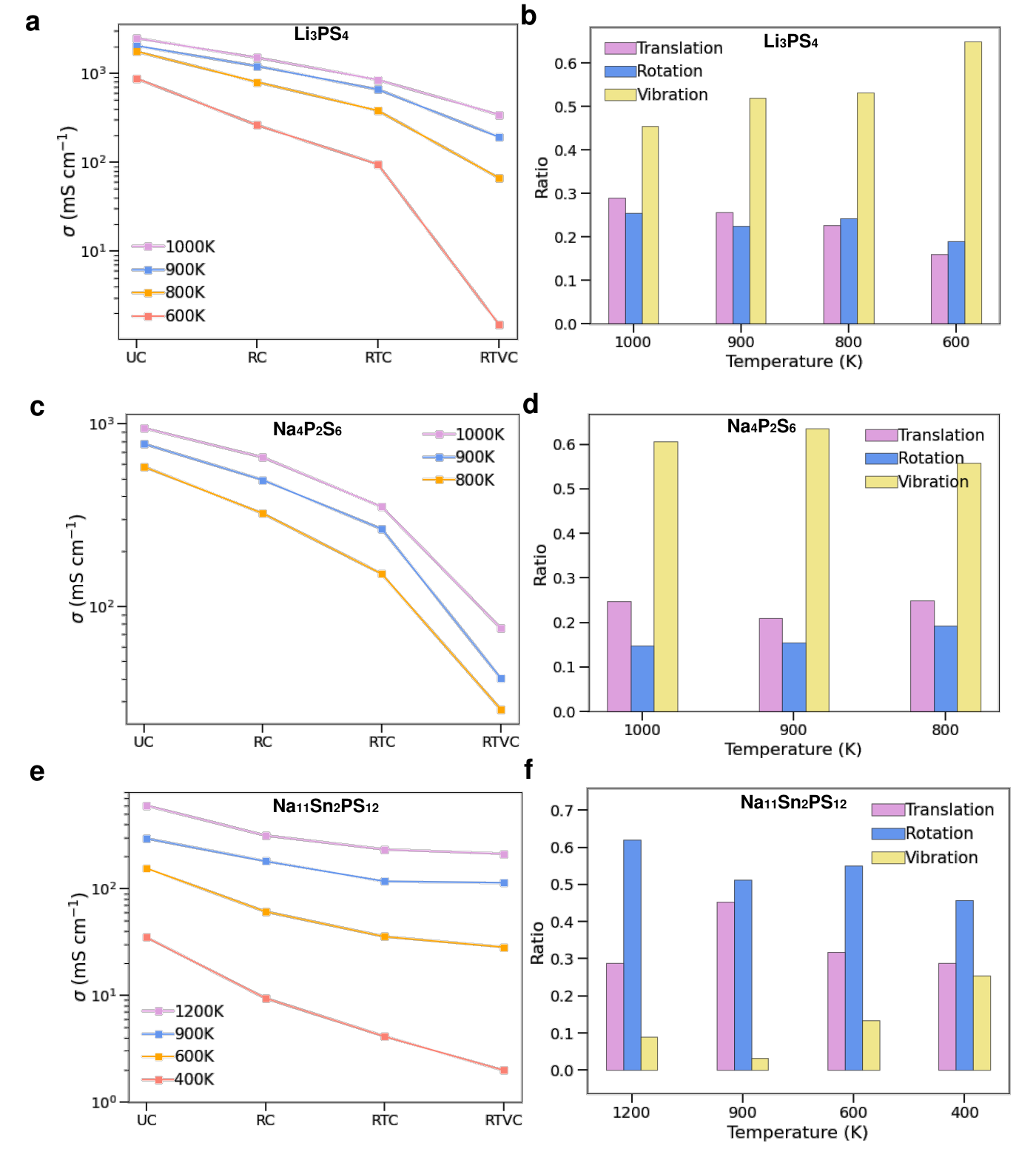}
\caption*{\textbf{Figure\:S12.}~Conductivity of (a) \ch{Li3PS4}, (c) \ch{Na4P2S6}, and (e) \ch{Na11Sn2PS12} under different constraints at various temperatures. The percentage ratio of the decrease in conductivity caused by constraining different anion motion modes of (b) \ch{Li3PS4}, (d) \ch{Na4P2S6}, and (f) \ch{Na11Sn2PS12}. In \ch{Li3PS4} and \ch{Na4P2S6}, anion vibration remains the dominant factor influencing conductivity. In \ch{Na11Sn2PS12}, both translation and rotation are dominant, while \ch{PS4^{3-}} vibration has a minimal effect.}% For \ch{Na2B10H10} and \ch{LiBH4}, where the rotation of \ch{B10H10^{2-}} and the translation of \ch{BH4^-} dominate, directly constraining these dominant motion modes leads to a drastic reduction in conductivity, causing it to fall below the measurable range. This results in the contribution ratio of each motion mode being undefined.}
\label{figS2}
\end{figure}

\begin{figure}[H]
\centering
\includegraphics[width=\textwidth]{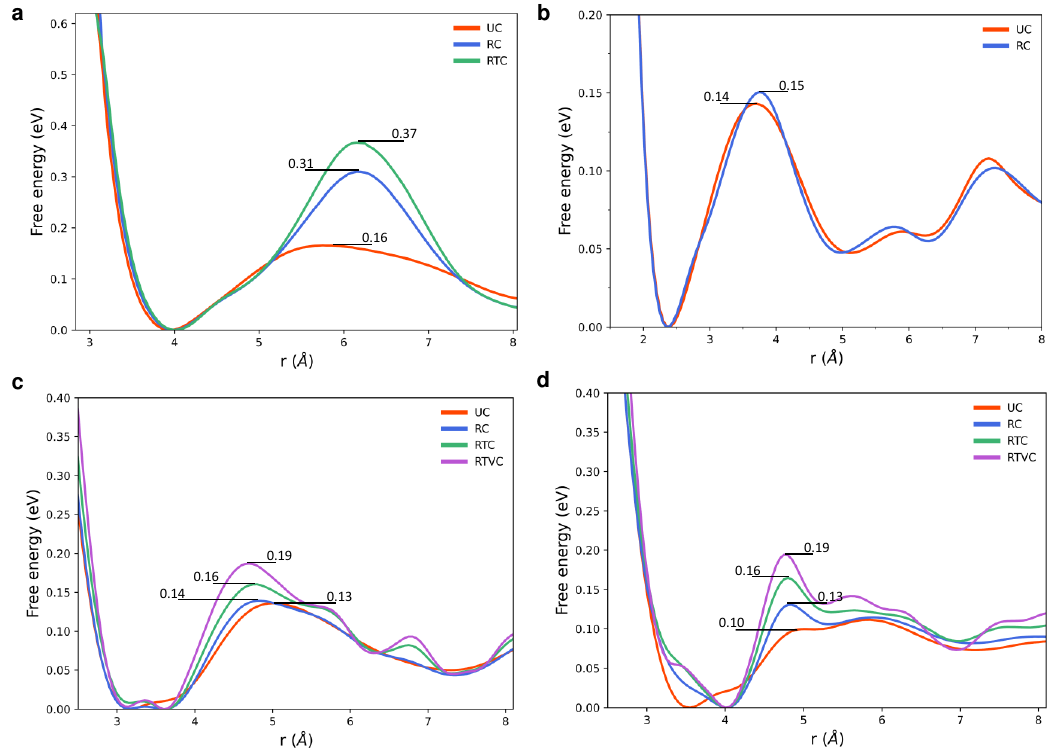}
\caption*{\textbf{Figure\:S13.}~Free energy as a function of distance between cations and anion center of mass under different constrained conditions in the (a) \ch{Na2B10H10} (750 K) (b) \ch{LiBH4} (550 K) (c) \ch{Li3PS4} (900 K), (d) \ch{Na4P2S6} (1000 K), and (e) \ch{Na11Sn2PS12} (900 K) systems.}
\label{figS9}
\end{figure}

\begin{figure}[H]
\centering
\includegraphics[width=\textwidth]{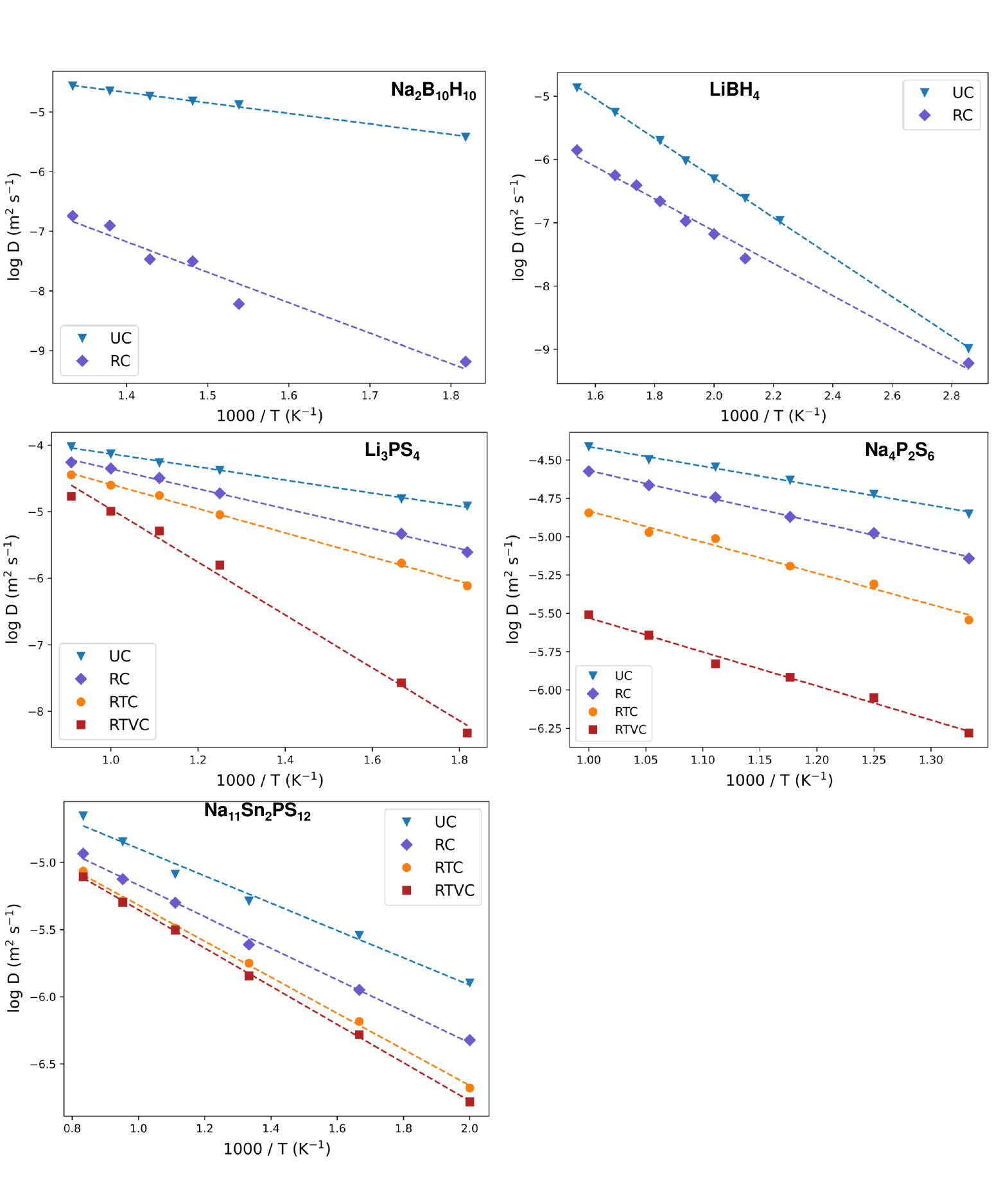}
\caption*{\textbf{Figure\:S14.}~Arrhenius plot of the calculated cation diffusion coefficient under different constraints.}
\label{figS12}
\end{figure}

\begin{figure}[H]
\centering
\includegraphics[width=\textwidth]{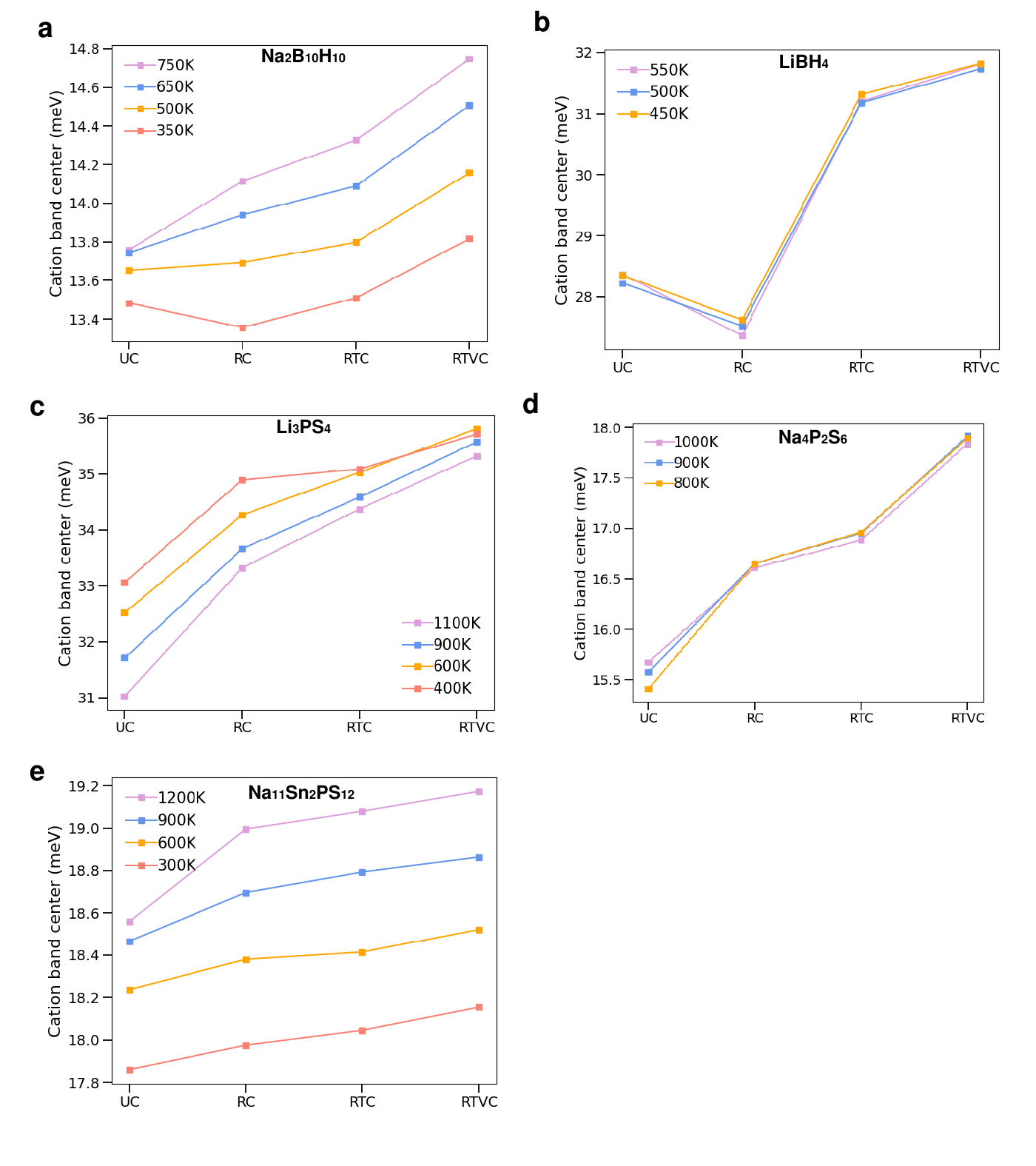}
\caption*{\textbf{Figure\:S15.}~Variation of the cation phonon band center under different constraints at various temperatures.}
\label{figS10}
\end{figure}

\begin{figure}[H]
\centering
\includegraphics[width=\textwidth]{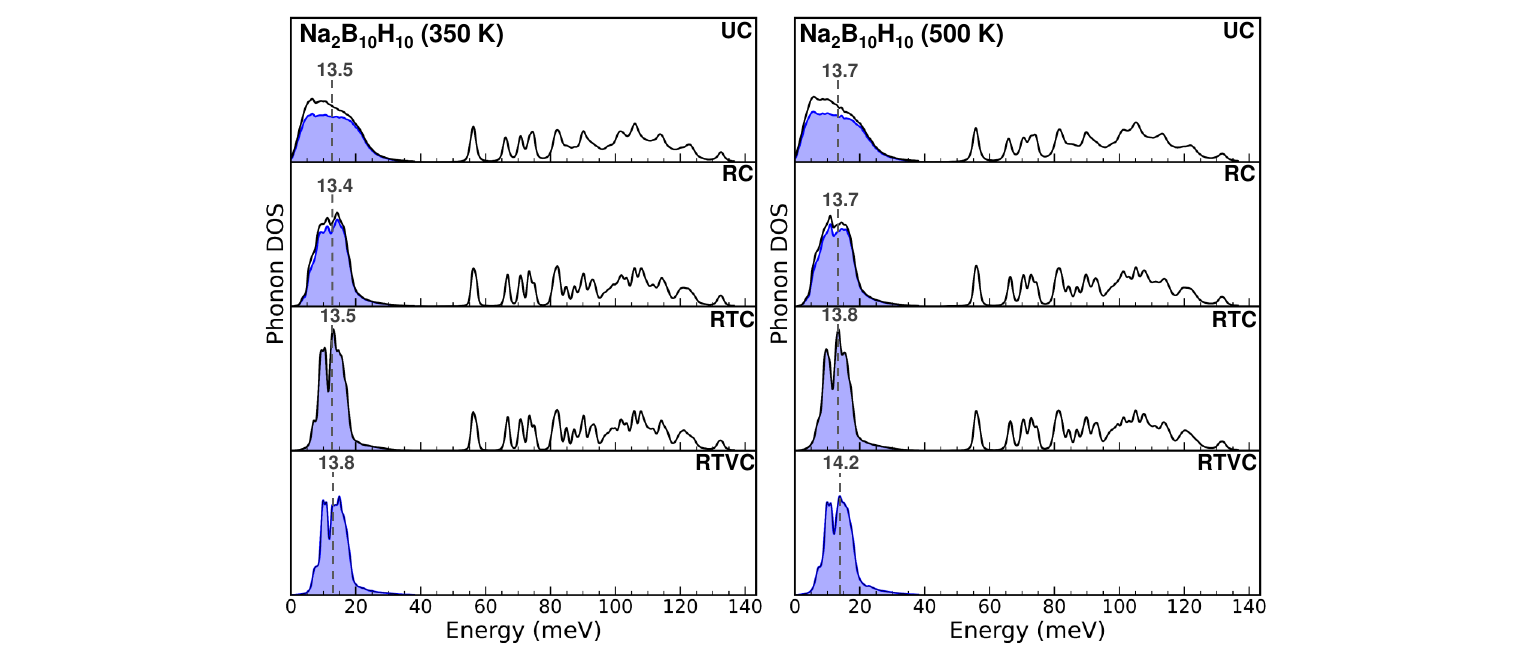}
\includegraphics[width=\textwidth]{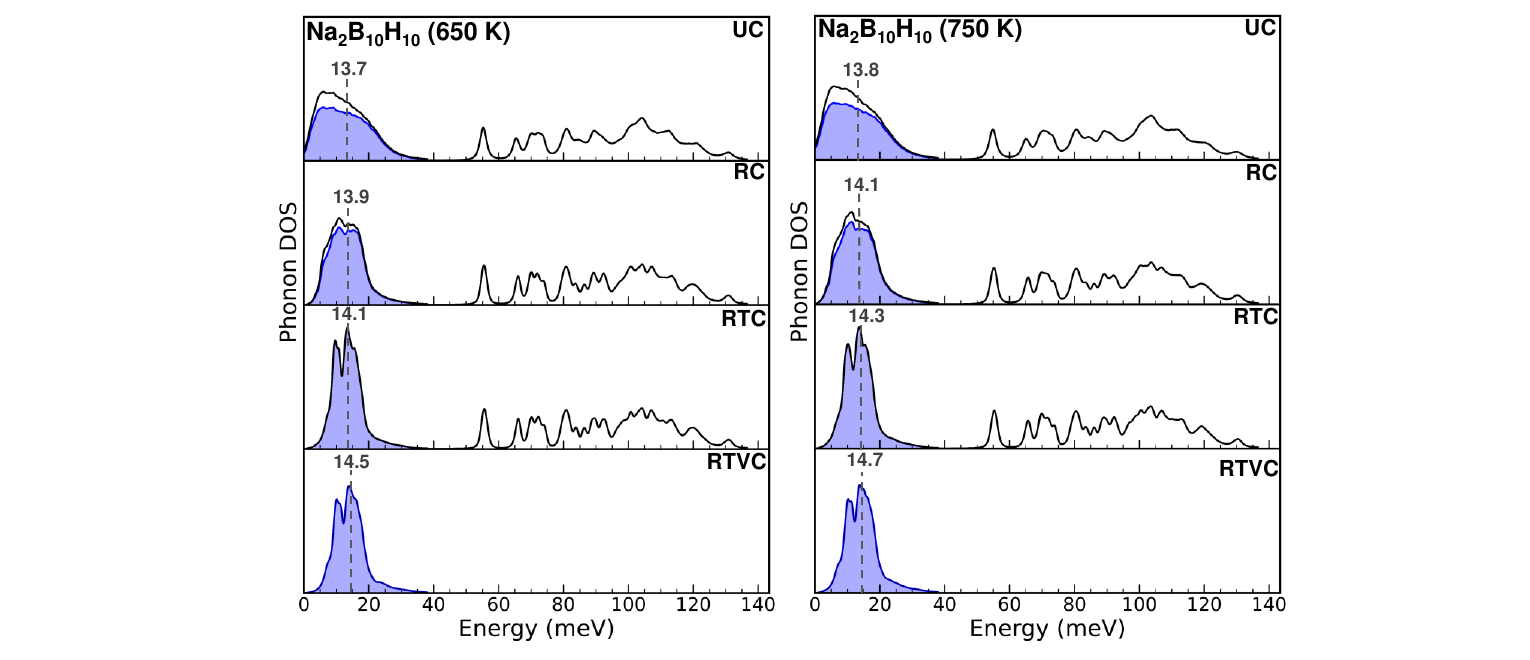}
\caption*{\textbf{Figure\:S16.}~Phonon density of states of \ch{Na2B10H10} under different constraints at various temperatures. The black lines represent the total phonon DOS, while the blue shaded areas correspond to the mobile cation-projected phonon DOS. The number and dashed line represent the cation band center. Higher frequency peaks are not plotted.}
\label{figS2}
\end{figure}

\begin{figure}[H]
\centering
\includegraphics[width=\textwidth]{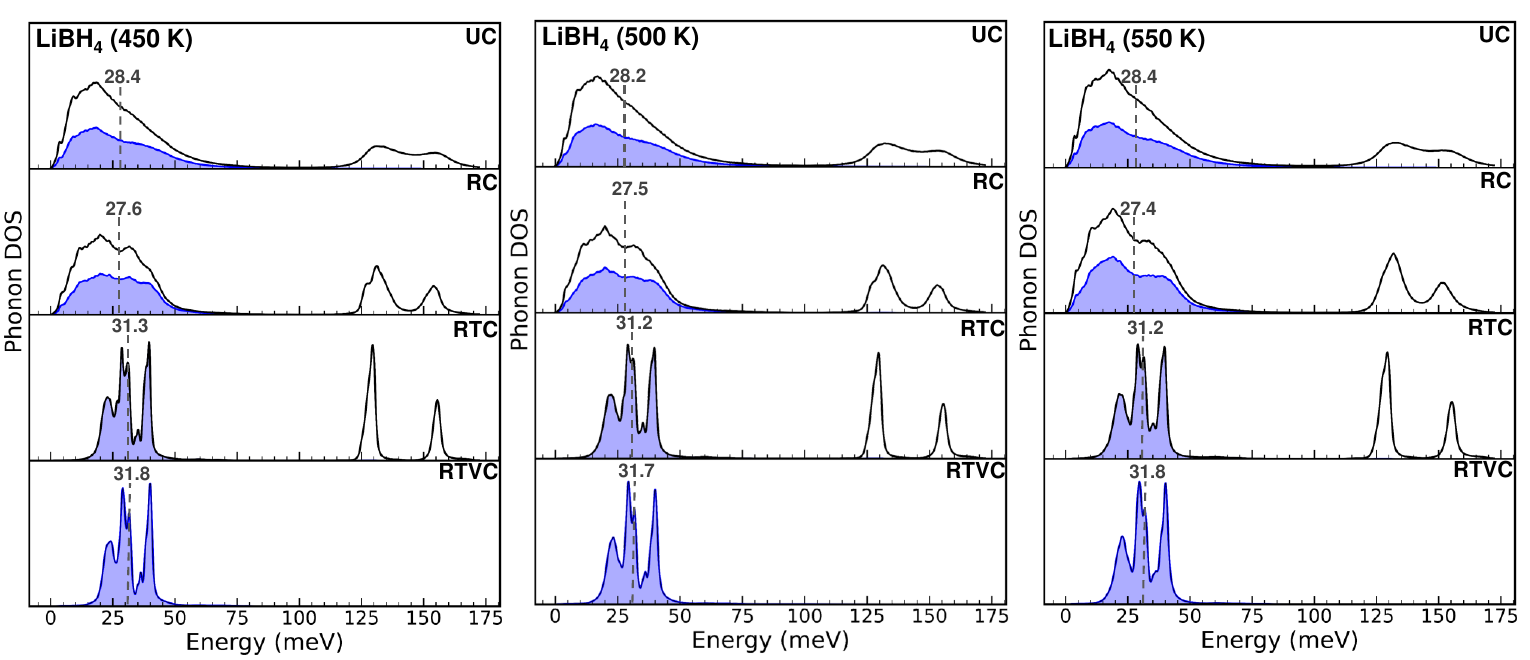}
\caption*{\textbf{Figure\:S17.}~phonon DOS of \ch{LiBH4}}
\label{figS2}
\end{figure}
\newpage
\begin{figure}[H]
\centering
\includegraphics[width=\textwidth]{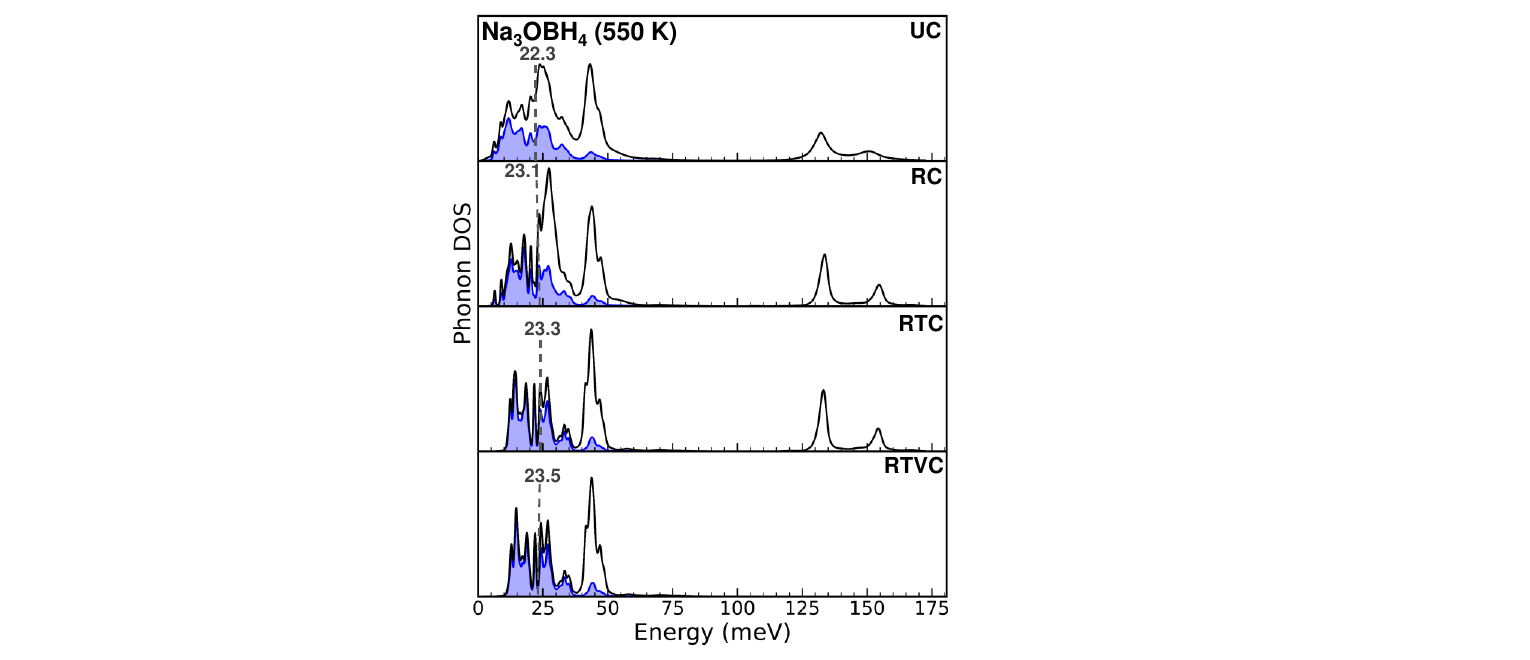}
\caption*{\textbf{Figure\:S18.}~phonon DOS of \ch{Na3OBH4}}
\label{figS2}
\end{figure}

\begin{figure}[H]
\centering
\includegraphics[width=\textwidth]{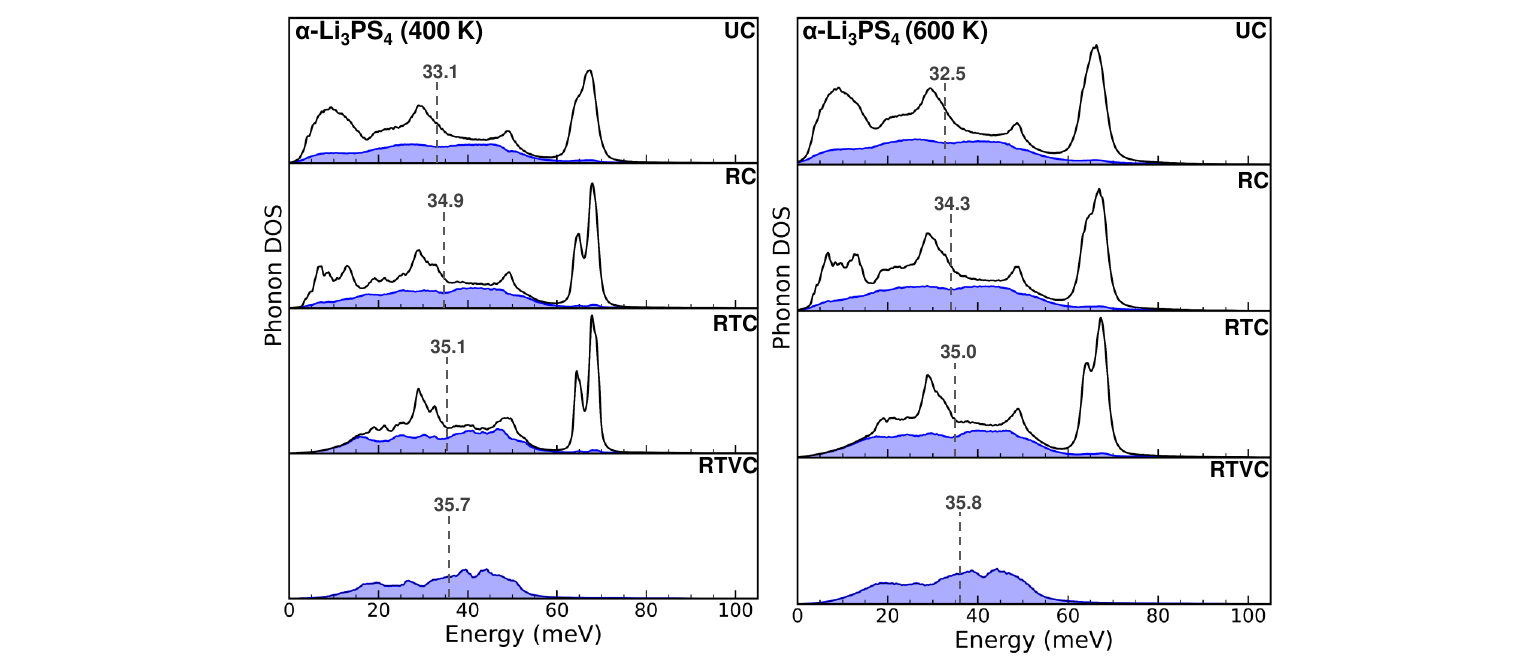}
\includegraphics[width=\textwidth]{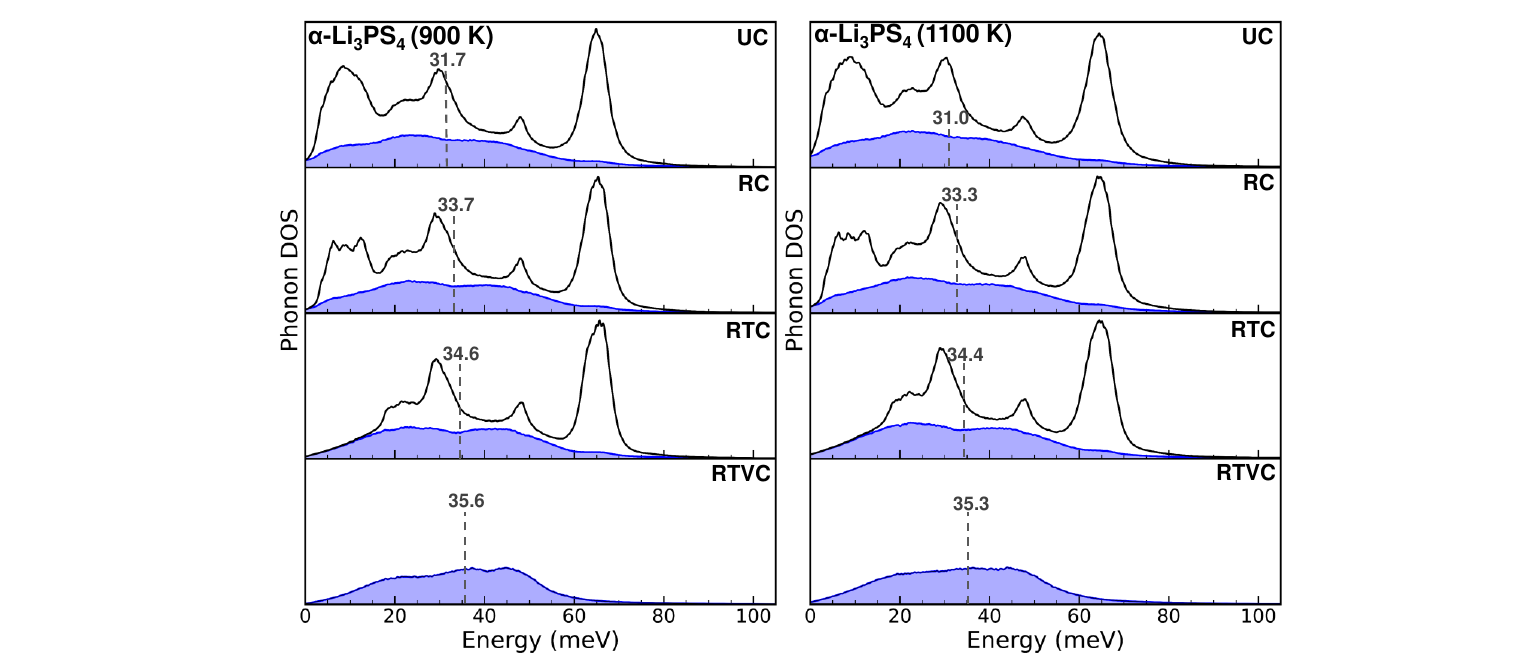}
\caption*{\textbf{Figure\:S19.}~phonon DOS of \ch{Li3PS4}}
\label{figS2}
\end{figure}

\begin{figure}[H]
\centering
\includegraphics[width=\textwidth]{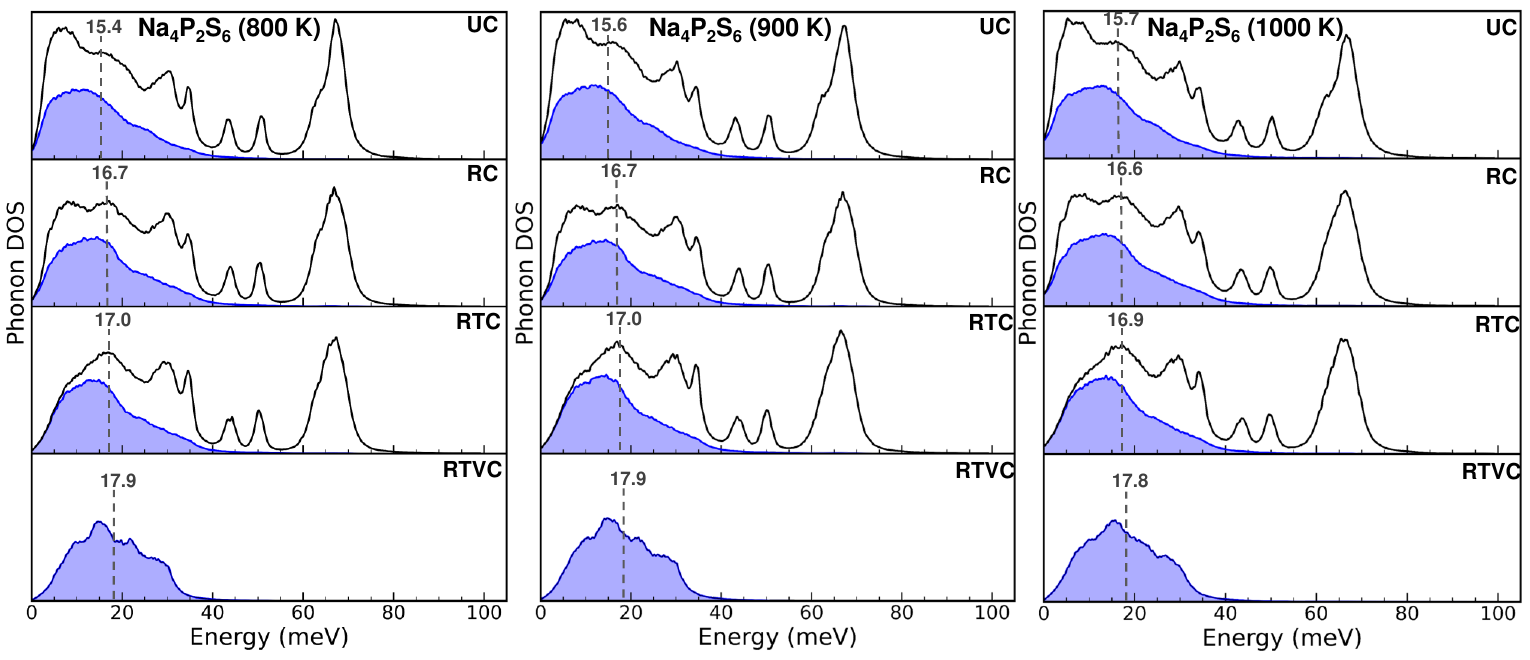}

\caption*{\textbf{Figure\:S20.}~phonon DOS of \ch{Na4P2S6}}
\label{figS2}
\end{figure}

\begin{figure}[H]
\centering
\includegraphics[width=\textwidth]{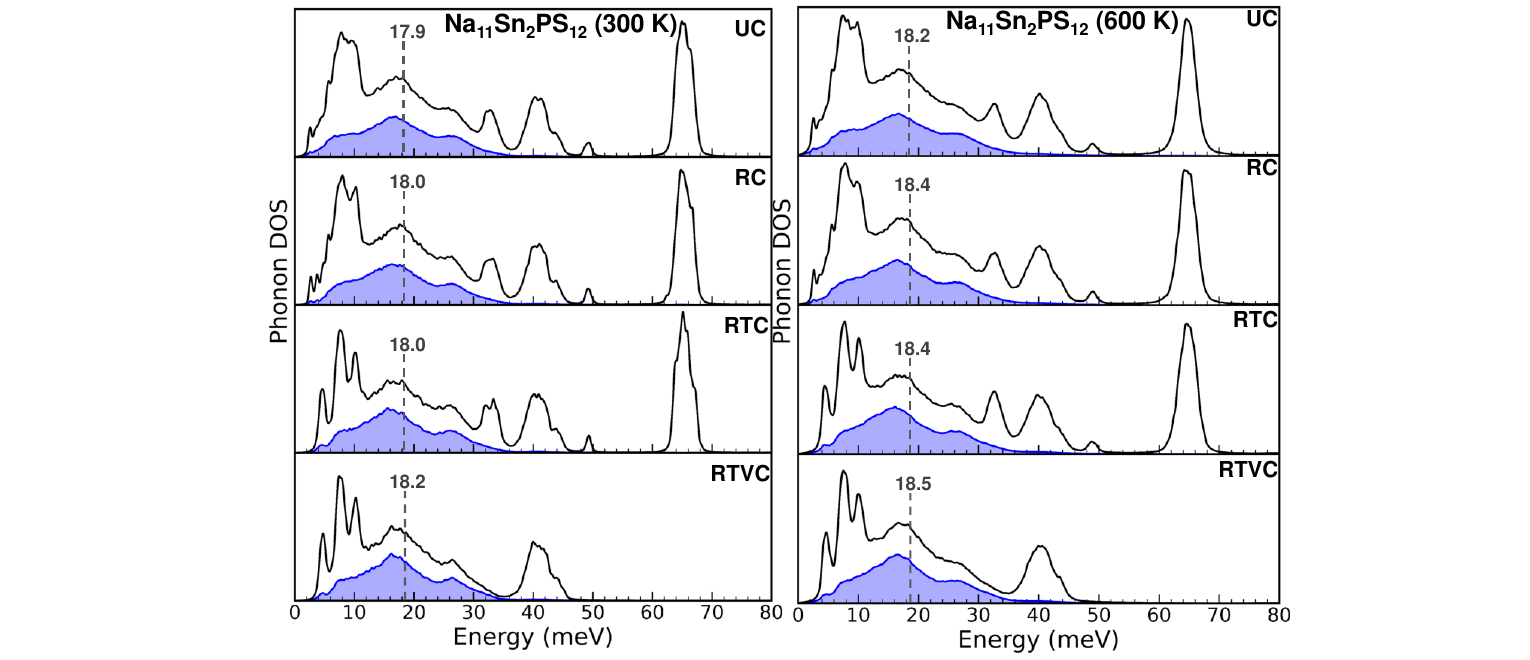}
\includegraphics[width=\textwidth]{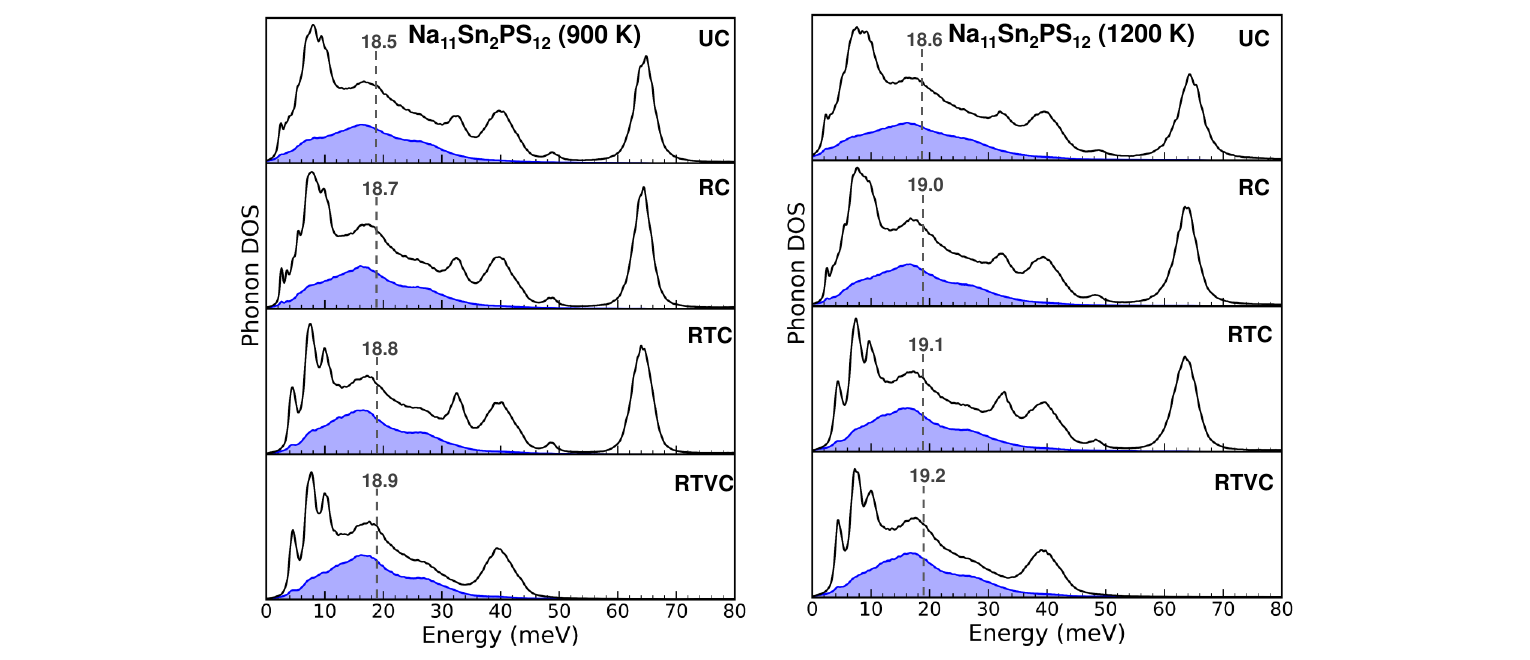}
\caption*{\textbf{Figure\:S21.}~phonon DOS of \ch{Na11Sn2PS12}}
\label{figS2}
\end{figure}

\begin{figure}[H]
\centering
\includegraphics[width=\textwidth]{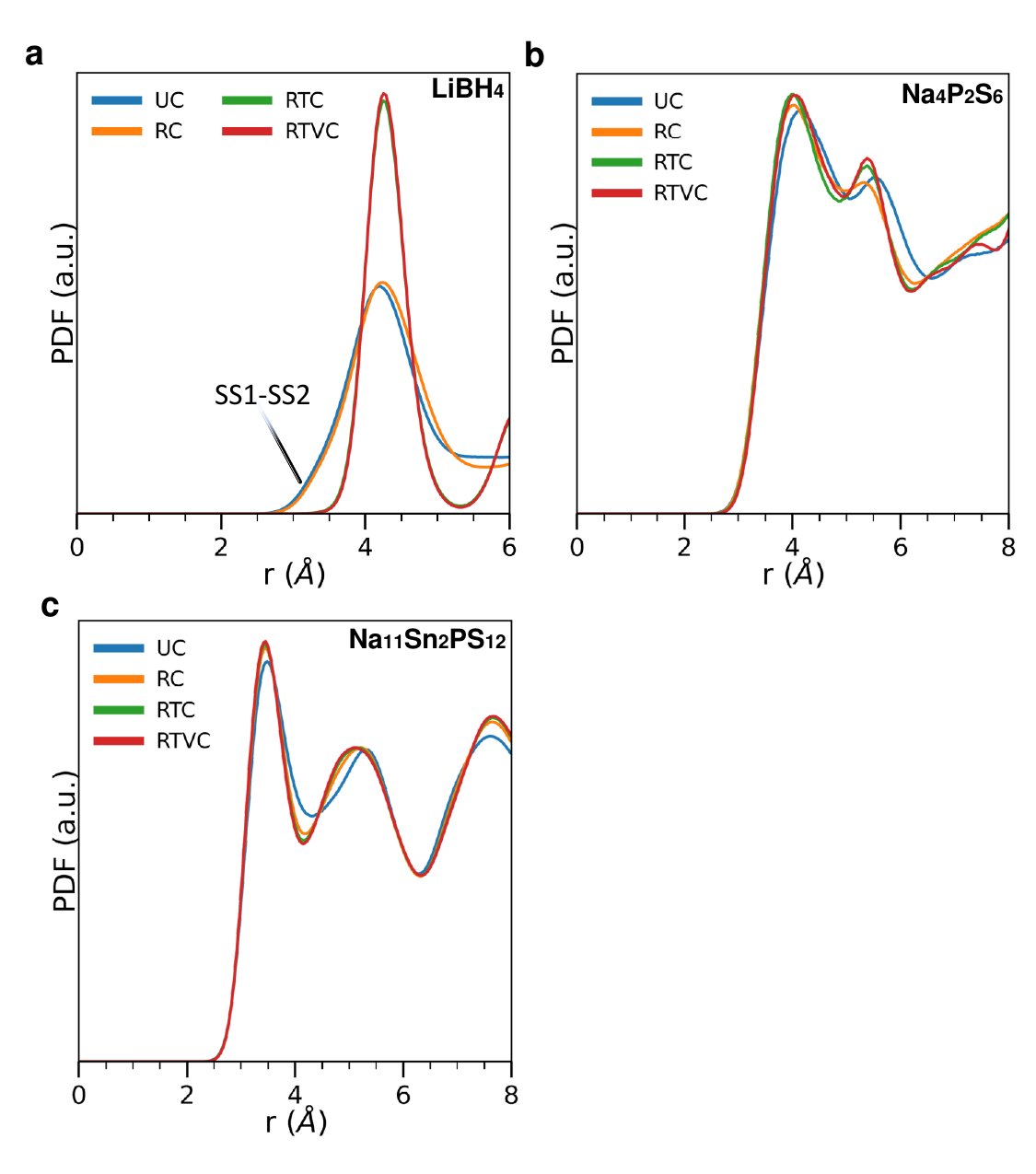}
\caption*{\textbf{Figure\:S22.}~Pair distribution function.}
\label{figS2}
\end{figure}

\begin{figure}[H]
\centering
\includegraphics[width=0.5\textwidth]{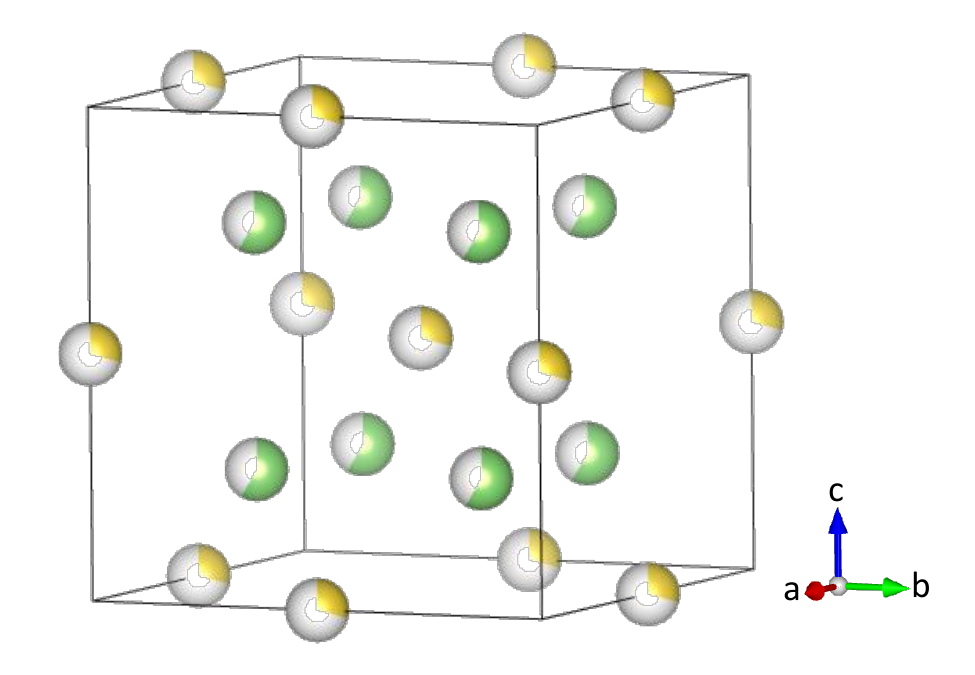}
\caption*{\textbf{Figure\:S23.}~Partial occupancies of \ch{Na^+} in \ch{Na2B10H10}. Only the partially occupied \ch{Na^+} sites are shown. Different colors represent different types of \ch{Na^+} occupied sites, with the proportion of each colored region to the white sphere corresponding to the partial occupancies.}
\label{figS2}
\end{figure}

%\begin{figure}[H]
%\centering
%\includegraphics[width=\textwidth]{pics/rotor/delta_linewidth.pdf}
%\caption*{\textbf{Figure\:S23.}~Decreases in linewidth are observed due to constraints on anion dynamics. Constraining any anion motion typically leads to a reduction in linewidth, with the dominant mode affecting the linewidth the most. To quantify this, we calculate the change in linewidth, by subtracting the linewidth of the RTVC from that of the UC. In materials like \ch{LiBH4}, \ch{Na2B10H10}, and %\ch{Li3PS4}, the broadening effect is much more pronounced, whereas in \ch{Na11Sn2PS12} and \ch{Na4P2S6}, the broadening of the cation phonon DOS due to anion dynamics is considerably less significant. The change in linewidth does not vary much with temperature.}
%\label{figS2}
%\end{figure}

%\begin{figure}[H]
%\centering
%\includegraphics[width=\textwidth]{pics/rotor/lw.pdf}
%\caption*{\textbf{Figure\:S24.}~Linewidth of five electrolytes under different constrained conditions at various temperatures.}
%\label{figS2}
%\end{figure}

%\begin{figure}[H]
%\centering
%\includegraphics[width=\textwidth]{pics/rotor/lw_ratio.pdf}
%\caption*{\textbf{Figure\:S25.}~The percentage ratio of change in linewidth due to each anion motion mode. The broadening of the cation phonon DOS due to anion dynamics in \ch{Na11Sn2PS12} and \ch{Na4P2S6} is much less pronounced, making it impractical to assess the contribution of each individual anion motion mode to the linewidth. Therefore, only electrolytes in which anion dynamics most significantly affect the linewidth are presented.}
%\label{figS2}
%\end{figure}

\begin{figure}[H]
\centering
\includegraphics[width=\textwidth]{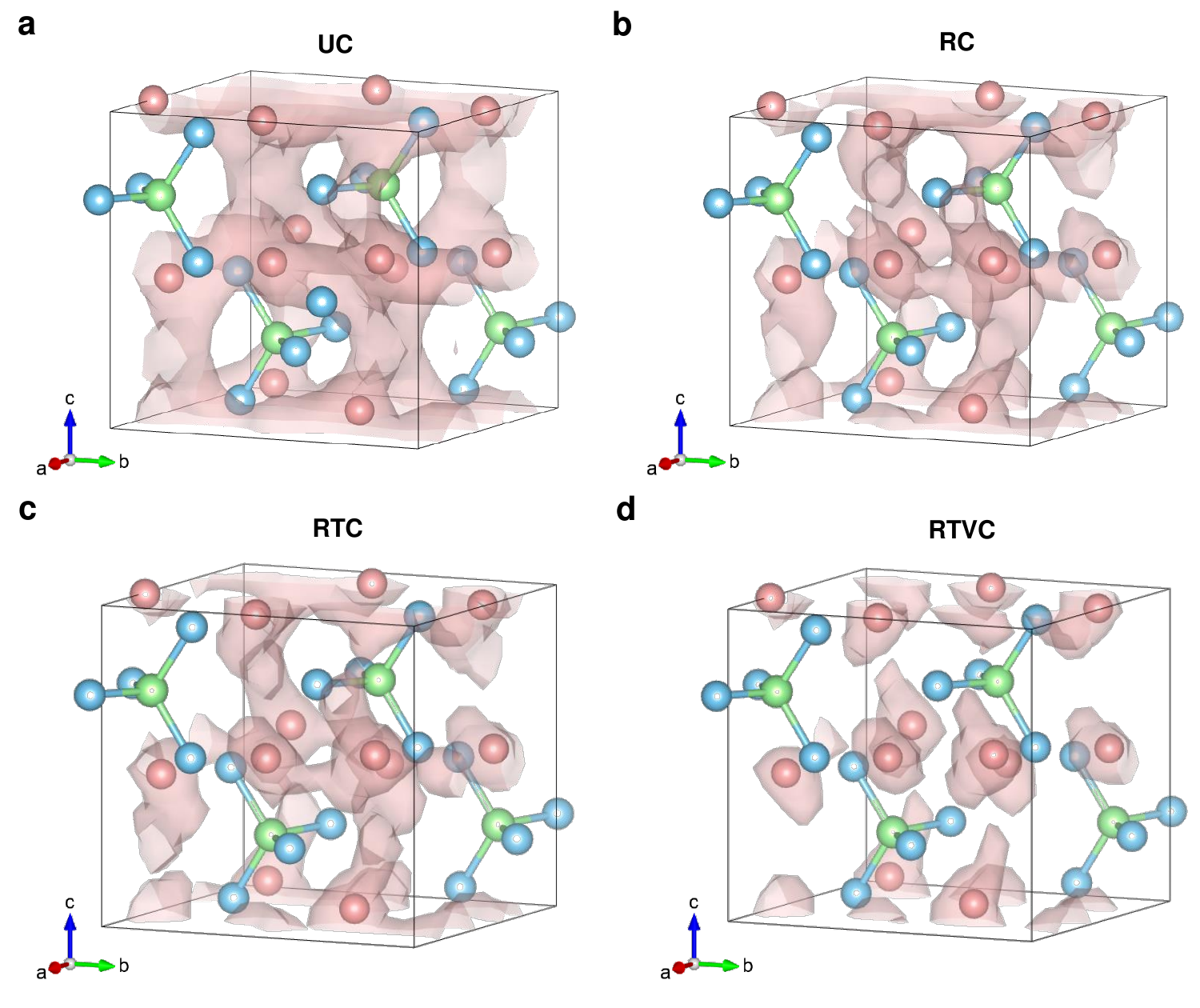}
\caption*{\textbf{Figure\:S24.}~\ch{Li^+} probability densities in \ch{Li3PS4} with an isosurface value of 0.001.}
\label{figS2}
\end{figure}

\begin{figure}[H]
\centering
\includegraphics[width=0.75\textwidth]{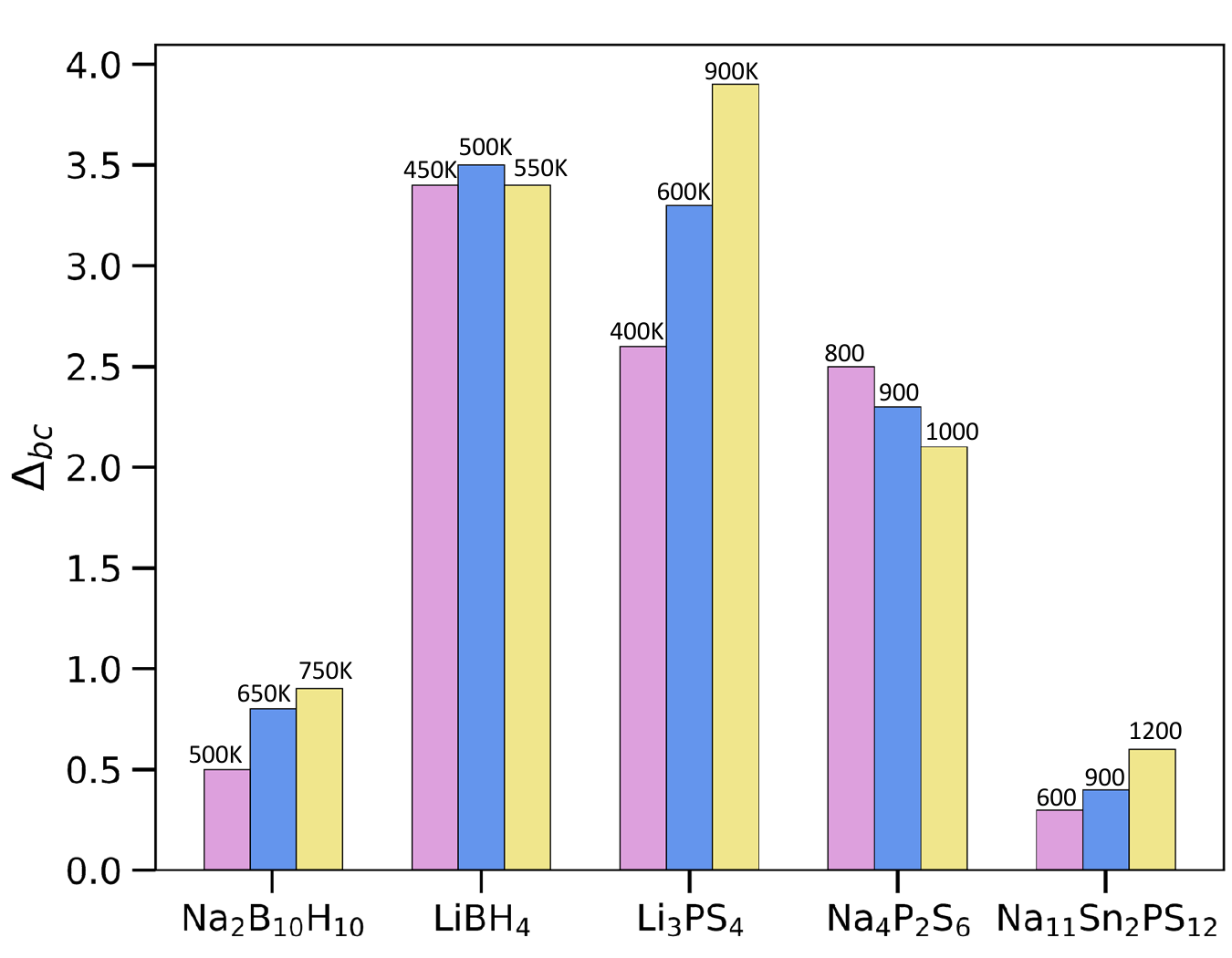}
\caption*{\textbf{Figure\:S25.}~Change in the cation phonon band center due to anion dynamics, calculated by subtracting the cation band center under RTVC from that under UC.}
\label{figS2}
\end{figure}

%\begin{figure}[H]
%\centering
%\includegraphics[width=\textwidth]{pics/rotor/disorder.pdf}
%\caption*{\textbf{Figure\:S23.}~Partial occupancies of \ch{Li^+} in \ch{Li3PS4}. The blue and green sphere are sulfur and phosphorus atoms, respectively. Different colors represent different types of \ch{Na^+} occupied sites, with the proportion of each colored region to the white sphere corresponding to the partial occupancies.}
%\label{figS2}
%\end{figure}

\FloatBarrier
%\bibliographystyle{plain}
%\subsection{References}




\bibliography{achemso-demo}